\documentclass[10pt]{article}
\usepackage[a4paper, left=1in, right=1in, top=1in, bottom=1in]{geometry}
\usepackage[affil-it]{authblk}
\usepackage{multirow}
\usepackage{graphicx}
\usepackage{amsmath}
\usepackage{xcolor}
\usepackage{arydshln}
\usepackage[ruled,vlined]{algorithm2e}
\usepackage[numbers,sort&compress]{natbib}
\usepackage[pdfborder={0 0 0}]{hyperref}
\usepackage[font=small,labelfont=bf]{caption}
\usepackage{caption}
\begin{document}
\title{\large\bf Dynamic Vaccine Prioritization via Non-Markovian Final-state Optimization}
\author[1,4]{Mi Feng}
\author[1,3]{Liang Tian\thanks{liangtian@hkbu.edu.hk}}
\author[1,2,3]{Changsong Zhou\thanks{cszhou@hkbu.edu.hk}}

\affil[1]{Department of Physics, Hong Kong Baptist University, Kowloon Tong, Hong Kong SAR 999077, China}
\affil[2]{Centre for Nonlinear Studies and Beijing-Hong Kong-Singapore Joint Centre for Nonlinear and Complex Systems (Hong Kong), Hong Kong Baptist University, Kowloon Tong, Hong Kong SAR 999077, China}
\affil[3]{Institute of Computational and Theoretical Studies, Hong Kong Baptist University, Kowloon, Hong Kong SAR 999077, China}
\affil[4]{Institute for Research and Continuing Education, Hong Kong Baptist University, Shenzhen, Guangdong, 518057, China}
\renewcommand\Authands{, }
\date{}
\maketitle


\begin{abstract}

	Effective vaccine prioritization is critical for epidemic control, yet real outbreaks exhibit memory effects that inflate state space and make long-term prediction and optimization challenging. 
	As a result, many strategies are tuned to short-term objectives and overlook how vaccinating certain individuals indirectly protects others.
	We develop a general age-stratified non-Markovian epidemic model that captures memory dynamics and accommodates diverse epidemic models within one framework via state aggregation. 
	Building on this, we map non-Markovian final states to an equivalent Markovian representation, enabling real-time fast direct prediction of long-term epidemic outcomes under vaccination.
	Leveraging this mapping, we design a dynamic prioritization strategy that continually allocates doses to minimize the predicted long-term final epidemic burden, 
	explicitly balancing indirect transmission blocking with the direct protection of important groups and outperforming static policies and those short-term heuristics that target only immediate direct effects.
	We further uncover the underlying mechanism that drives shifts in vaccine prioritization as the epidemic progresses and coverage accumulates, underscoring the importance of adaptive allocations. 
	This study renders long-term prediction tractable in systems with memory and provides actionable guidance for optimal vaccine deployment.

\end{abstract}

\section*{Introduction}
	
	Vaccination is among the most effective tools for curbing epidemic spread, and model-informed prioritization is critical for placing limited doses where they yield the greatest public-health benefit~\cite{anderson1991infectious,hethcote2000mathematics,keeling2008modeling,medlock2009optimizing}. 
	By linking quantitative predictions to policy objectives, such approaches provide a principled basis for comparing allocation schemes and adapting epidemic interventions as an outbreak evolves, a value underscored during the COVID-19 pandemic~\cite{bubar2021,matrajt2020,buckner2021,han2021time,yuan2024dynamic,matrajt2021optimizing,chen2022strategic}.

	However, many epidemics, including the COVID-19 pandemic, exhibit memory (non-Markovian) effects that complicate the optimization of long-term vaccination strategies.
	Such effects imply that the current transmission risk depends on the time since infection or vaccination, so event-time distributions deviate from the exponential law assumed in classical Markovian models~\cite{ganyani2020estimating,he2020temporal,to2020temporal,tian2021harnessing,ferretti2020quantifying,feng2019equivalence,feng2023validity,lin2020non,lin2024higher,starnini2017equivalence,kiss2015generalization,vanmieghem2013,vazquez2007}.
	For instance, generation intervals and infectiousness profiles are typically right-skewed and peak days after infection, a pattern documented repeatedly for COVID-19 disease~\cite{ganyani2020estimating,he2020temporal,to2020temporal,tian2021harnessing,ferretti2020quantifying}. 
	To capture these effects, diverse formulations such as delay-differential equations, age-of-infection integral kernels, and linear-chain stage expansions are widely used and well justified~\cite{feng2019equivalence,feng2023validity,lin2020non,lin2024higher,starnini2017equivalence,kiss2015generalization,vanmieghem2013,vazquez2007,hong2024overcoming,wearing2005appropriate,hurtado2021building,o2002tractable,hurtado2019generalizations,krylova2013effects}.
	While capturing memory effects makes epidemic models more realistic and general, it also increases the computational complexity of designing optimal intervention strategies, particularly for vaccination.
	In practice, the main bottleneck is the inability to perform direct long-term predictions under vaccination within an optimization loop:
	optimization must rely on repeated forward simulations to evaluate each candidate allocation strategy,
	which is computationally demanding when models must resolve history or infection-age structure with fine discretization, or when linear-chain expansions enlarge the explicit state dimension~\cite{hong2024overcoming,wearing2005appropriate,hurtado2021building,o2002tractable,hurtado2019generalizations,krylova2013effects}.
	The challenge is not how well these approaches can describe epidemics, but the lack of a fast route from current epidemic state to long-term outcomes that can be called inside an optimizer.
	
	These computational bottlenecks have direct policy consequences, constraining how vaccination strategies are designed and evaluated in practice. 
	To reduce computational cost, many vaccine allocation rules optimize short-term proxies, such as immediate incidence or near-term deaths, 
	thereby neglecting the long-term and indirect effects that vaccinating certain individuals confers on others (in some cases, this short-term focus may also reflect an effort to meet urgent near-term needs)~\cite{han2021time,medlock2009optimizing,fine2011herd,halloran1997study,fine2011herd}.
	Such rules evaluate only the immediate reduction that follows vaccination, but because the indirect effects of vaccination take time to emerge, they fail to capture how changes in one group’s susceptibility or infectiousness reshape the transmission network over time. 
	This narrow focus can overlook scenarios where prioritizing high-transmission groups ultimately prevents more cases and deaths than a strategy targeting only high-risk groups~\cite{medlock2009optimizing,bubar2021,matrajt2020,anderson1985vaccination}. 
	A useful framework must predict long-horizon outcomes efficiently while integrating both direct and indirect effects, enabling allocation strategies that are both adaptive and optimal over the course of an outbreak~\cite{hethcote2000mathematics,keeling2008modeling}.
	
	We address these limitations in three steps: (i) constructing a realistic and general epidemic modeling framework; 
	(ii) developing a theoretical method to predict long‑term outcomes under vaccination; 
	and (iii) leveraging this predictor to design a dynamic vaccination strategy that minimizes the long‑term epidemic burden.
	
	First, we introduce a general age-stratified non-Markovian model that captures memory dynamics. 
	Compared with classical Markovian formulations, our modeling approach offers two key advantages. 
	(i) Realism: its non-Markovian dynamics explicitly account for how event probabilities depend on the time elapsed since infection, rather than being memoryless~\cite{feng2023validity,ferretti2020quantifying}. 
	(ii) Generality: the non-Markovian formulation can unify diverse multi-stage models within a single, compact macro-state framework via state aggregation~\cite{diekmann2013mathematical,inaba2017age}. 
	As a result, intervention policies derived from our model (e.g., vaccination prioritization) are applicable across a broad class of mechanistic models, yielding recommendations that are robust to modeling choices.
	
	Second, building on this model, we develop a novel method to map a non-Markovian final state to an equivalent Markovian representation, 
	which yields a fast, direct predictor of long-term (final-state) epidemic burden under vaccination at any given decision point.
	For clarity, throughout this paper, we operationalize ``long-term'' as a final‑state quantity: the absorbing state where transmission has ceased and both direct and indirect vaccination effects have fully materialized.
	The key idea of predicting final state is to treat the decision point as a new initial point and then derive each individual's remaining transmission capacity from infection age (time since infection).
	This allows us to construct a Markovian surrogate system that shares the same final size from that point forward, even if the transient paths differ~\cite{feng2023validity,feng2019equivalence}.
	After this dimensionality reduction, non-Markovian long-term outcomes can be computed by solving the final state of the equivalent Markovian system, 
	thereby eliminating the need for long forward simulations~\cite{kermack1927contribution,andreasen2011final,feng2023validity}. 
	Consequently, candidate allocations within subsequent optimization loops can be evaluated instantly and the indirect effects of vaccination can be incorporated without increasing computational complexity.
	
	Finally, leveraging this predictor, we design a dynamic prioritization strategy that continuously allocates doses, 
	with each allocation optimized to minimize the predicted final epidemic burden, achieving a balance between indirect transmission blocking and direct protection of important groups.
	Across scenarios, our dynamic vaccination strategy consistently outperforms static allocations.
	The advantage of our long-term strategy over policies optimized for short-term goals depends on both the basic reproduction number ($R_0$) and the specific public-health objective.
	When $R_0$ is small and transmission is controllable, our strategy can outperform short-term planning by leveraging stronger indirect protection to block transmission; 
	and this advantage is especially pronounced for objectives such as minimizing deaths and years of life lost (YLL).
	Conversely, when $R_0$ is large and blocking becomes ineffective, our strategy can shift to prioritizing the direct protection of important groups; although this may resemble short-term planning, 
	it can still outperform such plans for certain objectives (e.g., YLL) due to its long-horizon perspective.
	
	A typical feature of dynamic vaccine prioritization is that, as vaccination proceeds over time, the target groups for vaccination switch at specific time points. 
	Revealing the mechanisms underlying this phenomenon not only deepens our understanding of how dynamic prioritization operates but also provides guidance for adjusting vaccination policies in real-world settings.
	To elucidate this mechanism, we quantify the time-varying marginal vaccination benefit (MVB) for each group, defined as the reduction in predicted final burden per additional dose.
	The optimal policy at each decision point allocates the next dose to the group with the highest MVB, and a change in which group occupies this top-MVB position signals a switch in vaccination priority.
	This further leads to two types of switches: when $R_0$ is low, the MVB curves of leading groups gradually converge over time, resulting in a partial switch, that is, a smooth rebalancing of allocation across groups; 
	when $R_0$ is high, the MVB curves intersect sharply, resulting in a full switch, namely a complete change in priority.
	
	Our study makes long-term prediction tractable in memory-dependent systems and provides practical guidance for vaccine deployment. 
	It addresses a central limitation of prior work by explicitly balancing direct and indirect effects, and offers implementable procedures for real-time prioritization. 
	This approach also provides a mechanistic rationale for dynamic allocation in epidemic control, guiding when and how vaccination should shift between population groups.

\section*{Results}
	This section outlines the core findings of our study. 
	First, in the subsection ``Model Framework'', we present the model construction, study objectives, and generality of the framework.
	Then, in the subsection ``Dynamic Vaccine Prioritization Based on Final-state Optimization'', we elaborate on the methodology for final-state prediction under vaccination, based on which we design the dynamic prioritization strategy.
	Next, the subsection ``Robustness from Dynamic Adaptation'' compares our approach with static strategies, demonstrating the robustness of dynamic allocation across diverse scenarios.
	Meanwhile, ``Robustness from Final-state Optimization'' evaluates our framework against dynamic strategies targeting short-term burdens, highlighting the advantages of optimizing for final-state outcomes.
	Furthermore, ``Mechanistic Insights into Vaccination Priority Switching'' analyzes the mechanism of vaccination switches during epidemic progression by introducing the concept of MVB.
	Finally, in the subsection ``Application: Dynamic Vaccine Prioritization in the COVID-19 Pandemic'', we use COVID-19 as a case study and draw on epidemiological data from multiple countries to demonstrate how our framework can be applied to real-world scenarios, 
	providing insights to inform vaccination policies in future pandemics.

	\subsection*{Model Framework}
	\begin{figure}[htbp]
		\centering
		\includegraphics[width=\linewidth]{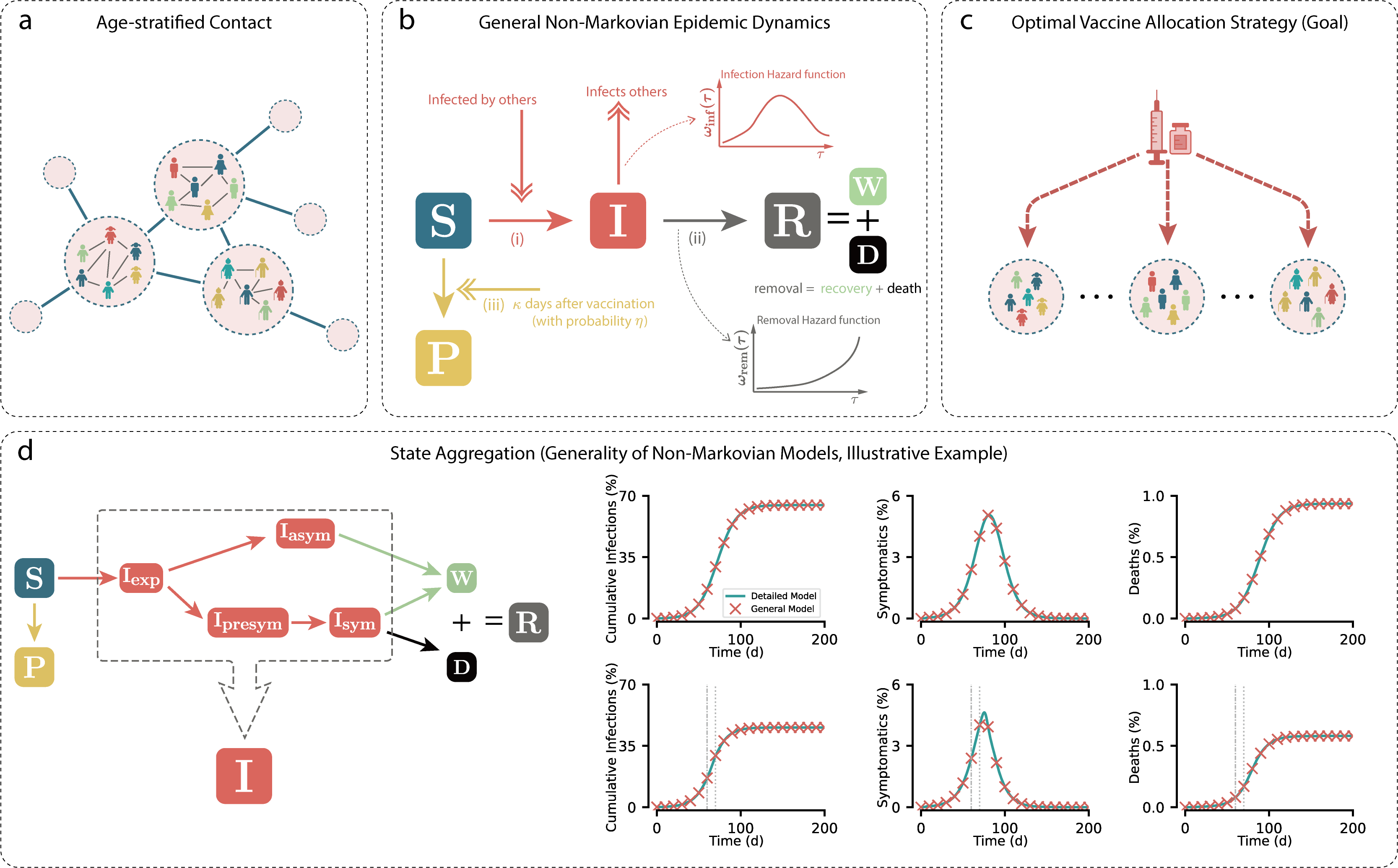}
		\caption{Model Construction. 
		\textbf{a} Age-stratified Contact. 
		The population is divided into subgroups according to their age. Contacts occur both within age groups and between different age groups. 
		\textbf{b} General Non-Markovian Epidemic Dynamics.
		Each individual is in one of four states: susceptible ($\textbf{S}$), infected ($\textbf{I}$), removed ($\textbf{R}$; recovery $\textbf{W}$ or death $\textbf{D}$), or protected ($\textbf{P}$). 
		Transitions occur via: (i) infection of susceptibles through contact with infectious individuals ($\textbf{S}\rightarrow\textbf{I}$); 
		(ii) removal of infected individuals by recovery or death ($\textbf{I}\rightarrow\textbf{R}$), with death governed by age-stratified infection fatality rates (IFRs); 
		and (iii) vaccination-induced protection ($\textbf{S}\rightarrow\textbf{P}$), occurring $\delta$ days after vaccination with probability $\eta$ (vaccine efficacy). 
		Time-varying infectiousness and removal are characterized by the hazard functions $\omega_{\mathrm{inf}}(\tau)$ and $\omega_{\mathrm{rem}}(\tau)$, respectively. 
		\textbf{c} Optimal Vaccine Allocation Strategy (Goal). 
		Based on the model, this study aims to develop an optimal, model-informed strategy for dynamically allocating vaccines among different age groups over time under limited supply, 
		with the objective of minimizing the overall epidemic burden. 
		\textbf{d} State Aggregation (Generality of Non-Markovian Models, Illustrative Example). 
		As an illustrative example of the generality of non-Markovian models, 
		a detailed model with exposed ($\textbf{I}_{\mathrm{exp}}$), asymptomatic ($\textbf{I}_{\mathrm{asym}}$), presymptomatic ($\textbf{I}_{\mathrm{presym}}$), 
		and symptomatic ($\textbf{I}_{\mathrm{sym}}$) substates is aggregated to a single infected class $\textbf{I}$. 
		Simulations (blue: detailed; red $\times$: general) show show perfect agreement in the time evolution of cumulative infections, symptomatic cases and deaths 
		without vaccination (upper panels) and with vaccination (lower panels). 
		In the vaccination scenario, an idealized one-time campaign vaccinates 30\% of the total population; 
		the dash-dotted and dotted lines indicate the vaccination time and the onset of vaccine-induced protection, respectively.}
		\label{fig:framework}
	\end{figure}
	
	As shown in Fig.~\ref{fig:framework}a, we partition the population into age-stratified subgroups; the age composition is given by the vector $\boldsymbol{\rho}$. 
	Contacts occur both within groups and between groups. Patterns of interaction are represented by the network structure, 
	and contact intensities are encoded in the contact matrix $\mathbf{A}$. Within this age-structured population, the disease can spread among individuals.
	
	As shown in Fig.~\ref{fig:framework}b, each individual occupies one of four epidemiological states: 
	susceptible ($\textbf{S}$), infected ($\textbf{I}$), removed ($\textbf{R}$; comprising recovery $\textbf{W}$ and death $\textbf{D}$), or protected ($\textbf{P}$). 
	Individuals transition between states via: (i) infection of the susceptible population through contact with infectious individuals ($\textbf{S}\rightarrow\textbf{I}$); 
	(ii) removal of infected individuals from transmission due to recovery or death ($\textbf{I}\rightarrow\textbf{R}$; removed individuals do not become infected again), 
	and (iii) vaccination, whereby susceptibles become protected a fixed number of days after vaccination ($\textbf{S}\rightarrow\textbf{P}$; for simplicity, we assume that protected individuals acquire permanent immunity against infection).
	This model extends a non-Markovian susceptible-infected-removed (SIR) framework by incorporating vaccine-induced protection, yielding a non-Markovian susceptible-infected-removed-protected (SIRP) model.
	Our model is closely related to formulations sometimes termed susceptible-infected-recovered-vaccinated-deceased, SIRVD, 
	in which the removed state is explicitly partitioned into recovered and deceased and the vaccinated state denotes vaccine-induced protection~\cite{schlickeiser2024mathematics,liao2021sirvd}.
	
	The model is non-Markovian because an infected individual’s infection and removal probabilities depend on the time since infection, referred to as the infection age $\tau$. 
	Time-dependent infectiousness of one individual is described by the infection hazard function $\omega_{\mathrm{inf}}(\tau)$, so that the probability of generating an infection in $[\tau,\tau+d\tau)$ is $\omega_{\mathrm{inf}}(\tau)d\tau$. 
	Time-dependent removal is governed by the removal hazard function $\omega_{\mathrm{rem}}(\tau)$ and the age-stratified infection fatality rates (IFRs), represented by the vector $\boldsymbol{\epsilon}$.
	The probability of removal within the interval $[\tau,\tau+d\tau)$ is $\omega_{\mathrm{rem}}(\tau)d\tau$. Conditional on removal, 
	individuals die with probabilities determined by their age-stratified component of $\boldsymbol{\epsilon}$; otherwise, they recover.
	Because $\omega_{\mathrm{inf}}(\tau)$ and $\omega_{\mathrm{rem}}(\tau)$ vary with $\tau$, the dynamics depend on when individuals were infected, imparting memory to the process. 
	The non-Markovian nature can also be characterized by the infection and removal time distributions, $\psi_{\mathrm{inf}}(\tau)$ and $\psi_{\mathrm{rem}}(\tau)$, 
	which can be determined by $\omega_{\mathrm{inf}}(\tau)$ and $\omega_{\mathrm{rem}}(\tau)$, and are generally non-exponential when these hazards are time-varying (see Methods for detailed definitions and derivations).
	Additionally, in this study we model the post-vaccination transition $\textbf{S}\rightarrow\textbf{P}$ with a fixed-delay kernel: 
	a vaccinated susceptible remains in the susceptible class during the $\delta$-day delay and can still be infected in that period; 
	if uninfected after $\delta$ days, the individual transitions to $\textbf{P}$ with probability $\eta$, where $\eta$ denotes the vaccine efficacy.
	Based on this framework, our goal is to design an optimal, model-informed vaccination strategy that dynamically allocates vaccines across age groups over time under limited supply, in order to minimize the overall epidemic burden, as illustrated in Fig.~\ref{fig:framework}c.
	For a detailed account of how our model's dynamic equations are established and how vaccination affects the dynamic variables, please refer to the Methods section.
	
	Before detailing how the model informs the optimal vaccination strategy, we first show that our formulation is sufficiently general. 
	As shown in Fig.~\ref{fig:framework}d, we consider a more detailed compartmentalization (hereafter the detailed model, in contrast to our general model). 
	In the detailed model, an infected individual first enters an exposed state ($\textbf{I}_{\mathrm{exp}}$) without symptoms and infectiousness. 
	From this state, the individual can transition to either an asymptomatic state ($\textbf{I}_{\mathrm{asym}}$) or a presymptomatic state ($\textbf{I}_{\mathrm{presym}}$). 
	Asymptomatic individuals are infectious but remain symptom-free throughout their infectious period, ultimately recovering. 
	Presymptomatic individuals are also infectious without symptoms but will progress to a symptomatic state ($\textbf{I}_{\mathrm{sym}}$), which remains infectious with symptoms and will eventually either recover or die.
	In this detailed model, each state is characterized by distinct infectiousness and transition rates, both dependent on the time since entering that state. 
	Through state aggregation, these four infected substates can be combined into a single state ($\textbf{I}$), yielding the general model described in Fig.~\ref{fig:framework}b. 
	The corresponding infection hazard function $\omega_{\mathrm{inf}}(\tau)$ and removal hazard function $\omega_{\mathrm{rem}}(\tau)$ can be derived from the parameters of the detailed model;
	meanwhile, the time-evolution curves of each substate in the detailed model can be reconstructed from the general model (see Supplementary Note 1 for details on derivation and reconstruction).
	The six panels on the right of Fig.~\ref{fig:framework}d compare simulation results of the detailed model (blue curves) and the general model (red $\times$ markers) for cumulative infections, symptomatic cases, and deaths, 
	both without vaccination (upper panels) and with vaccination (lower panels). The perfect agreement between the two models validates the theoretical framework and demonstrates the generality of our model.

	\subsection*{Dynamic Vaccine Prioritization Based on Final-state Optimization}
	\begin{figure}[htbp]
		\centering
		\includegraphics[width=\linewidth]{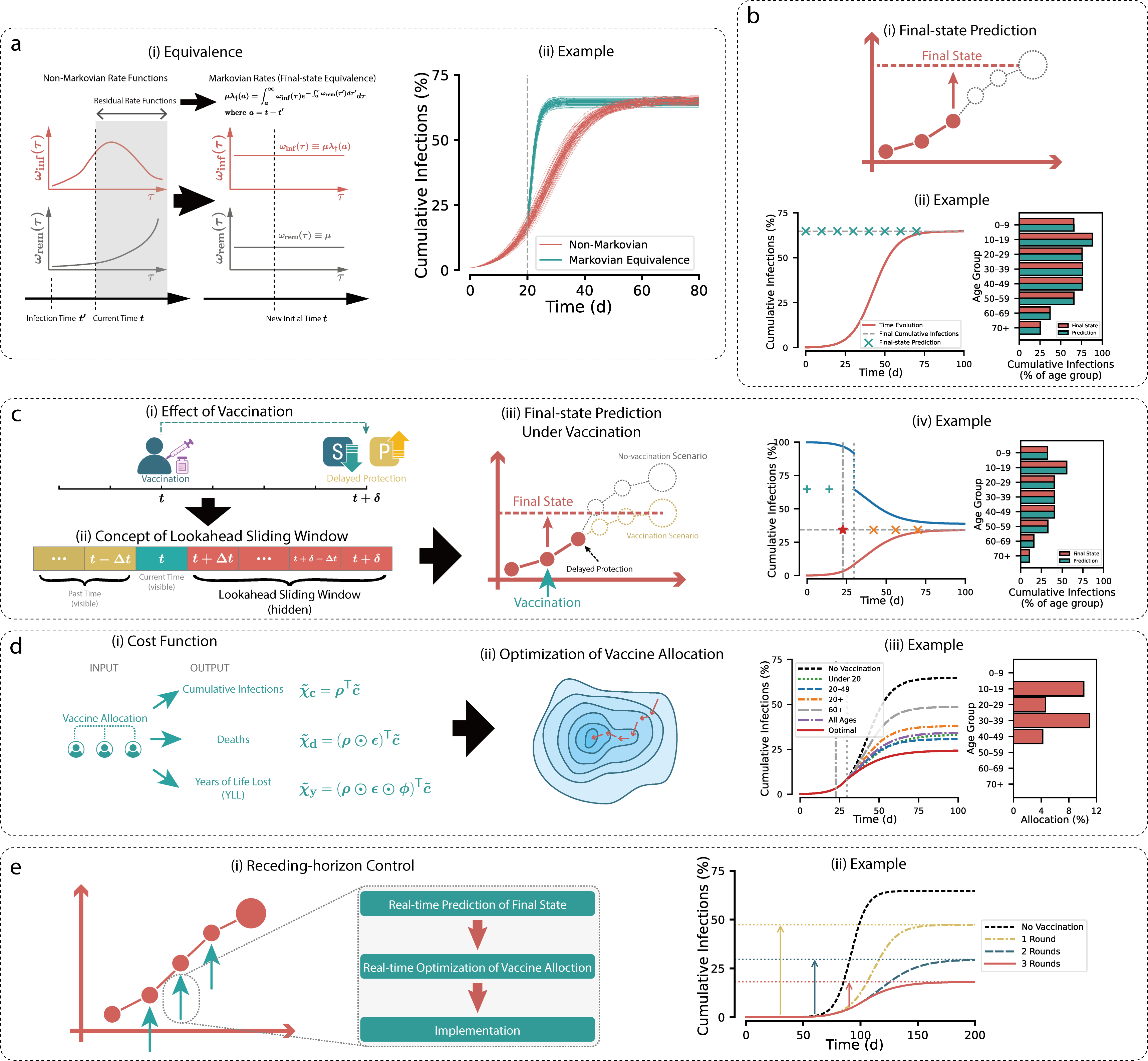}
		\caption{Dynamic vaccine prioritization via final-state Optimization.
			\textbf{a} Final-state equivalence. 
			(i): At an arbitrary time $t$, for an individual infected at $t'$, only hazards beyond the infection age $a = t-t'$ influence subsequent transmission, thereby defining residual effective infection rate $\lambda_{\dagger}(a)$. 
			Replacing hazard functions with $\omega_{\mathrm{inf}}(\tau)\equiv \mu\lambda_{\dagger}(a)$ and $\omega_{\mathrm{rem}}(\tau)\equiv \mu$ ($\mu>0$) yields a Markovian surrogate with the same final state.
			(ii): In $100$ Monte Carlo simulations (red curves), switching to the Markovian surrogate at an intermediate time $t$ yields trajectories (blue curves) that converge to identical final cumulative infections, confirming final-state equivalence.
			\textbf{b} Final-state prediction without vaccination. 
			(i): By solving the Markovian surrogate, the equivalence enables final-state prediction at any time without vaccination.
			(ii): Predictions made at distinct times (blue $\times$) match the final value (horizontal dashed line) of the cumulative-infection curve (red curve), including at the age-group level (bars).
			\textbf{c} Final-state prediction under vaccination. 
			(i): Vaccination reduces susceptibles, but protection is realized after a fixed $\delta$-day delay, complicating analysis.
			(ii): A lookahead sliding window precomputes $\delta$-ahead states, making protection realized at $t+\delta$ available for prediction already at vaccination time $t$.
			(iii): With this window, final state is predictable at any time under vaccination.
			(iv): Predictions at and after vaccination (blue $+$ before, red $\star$ at, orange $\times$ after vaccination) align with the final value (horizontal dashed line) of the cumulative infection curve (red curve) under vaccination, including at the age-group level (bars); 
			the blue curve illustrates the decline of susceptible population  over time and the abrupt reduction due to the delayed protection (dotted vertical line) following the vaccination (dash-dotted vertical line).
			\textbf{d} Vaccine Allocation Optimization.
			(i): Age-stratified final-state infections allow computation of total cumulative infections, deaths, and YLL, defining a cost function maping candidate vaccine allocation to final epidemic burden.
			(ii): The allocation that minimizes the predicted final epidemic burden is solved using Sequential Least Squares Programming (SLSQP) optimization algorithm.
			(iii): The optimal strategy outperforms five empirical reference strategies.
			\textbf{e} Dynamic Vaccine Prioritization.
			(i): At each vaccination time, we determine and administer the vaccine allocation that minimizes the predicted final epidemic burden at that time, and then repeat the process at the next decision point.
			(ii): After three rounds, the red curve shows cumulative infections; yellow, blue, and red arrows mark final‑state predictions at the 1st, 2nd, and 3rd rounds, 
			each using only information available up to that time and equaling the realized final size after one (yellow dash‑dotted), two (blue dashed), and three (red solid) rounds.}
		\label{fig:strategy}
	\end{figure}
	In this subsection, we detail the workflow for developing our dynamic vaccine prioritization based on minimizing the real-time predicted final states (Fig.~\ref{fig:strategy}). 
	The pipeline proceeds from a final-state equivalence, to fast prediction without and with vaccination, to optimization of allocations, and finally to dynamic vaccine prioritization.
	
	We first establish a final-state equivalence between non-Markovian and Markovian formulations that can be invoked at an arbitrary intermediate time (Fig.~\ref{fig:strategy}a). 
	At a chosen time $t$, each infected individual with its infection age $a=t-t'$ ($t'$ denotes its infection time) is assigned a residual effective infection rate $\lambda_{\dagger}(a)$. 
	From that time onward, replacing the original time-varying hazard functions by constants $\omega_{\mathrm{inf}}(\tau)\equiv \mu\lambda_{\dagger}(a)$ and $\omega_{\mathrm{rem}}(\tau)\equiv \mu$ (with $\mu>0$ arbitrary) 
	yields a Markovian surrogate that reaches the same final state. 
	By ``constant'' we mean fixed within the surrogate process after replacement, with values determined by each individual’s infection age $a$ at the time of replacement (for detailed analysis, see Methods; for derivations, see Supplementary Note 2).
	For example, across 100 independent Monte Carlo simulations of a non-Markovian outbreak, replacing the process at an arbitary intermediate time with the final-state equivalent Markovian surrogate produces convergence to the same final size, 
	which confirms the equivalence.
	
%
%
%
%
	This equivalence enables the prediction of the final epidemic state at any given time point in a scenario without vaccination (Fig.~\ref{fig:strategy}b).
	At any time during an outbreak we can transform the non-Markovian parameters to those of a Markovian process that is equivalent with respect to the final state, thereby reducing final-state prediction to a lower-dimensional problem. 
	We then solve the Markovian final state via the Kermack–McKendrick final-size relation (which, in the Markovian framework, determines the epidemic final state), 
	thereby obtaining a direct final-state prediction for the original non-Markovian dynamics at any chosen time (for detailed equations, see Methods; for derivations, see Supplementary Note 2)~\cite{kermack1927contribution,andreasen2011final}.
	In the example of Fig.~\ref{fig:strategy}b, predictions of the final cumulative infections made at different time points coincide with the ultimate value of the cumulative-infection curve, including at the age-group level. 
	This agreement supports the theory and enables direct prediction of the final state without simulating the entire transmission process.
	
	To account for vaccination, we use a lookahead sliding window that captures the $\delta$-day delay in protection and enables accurate final-state prediction under vaccination (Fig.~\ref{fig:strategy}c).
	In details, incorporating vaccination introduces a fixed protection delay of $\delta$ days, so the reduction in susceptibles is not immediate. 
	Consequently, a naive approach would simulate the dynamics forward to $t+\delta$ after each vaccination event before predicting the final state, which increases computational cost and complicates downstream optimization. 
	To address this, we introduce a lookahead sliding window that provides immediate access to the state at $t+\delta$ following a vaccination at time $t$, thereby enabling final-state prediction under delayed protection 
	(For detailed analyses, see Methods; for the full algorithm of epidemic simulation with a lookahead sliding window, see Supplementary Note 3). 
	As an example, predictions of the final cumulative infections made at the time of vaccination and at subsequent post-vaccination times coincide with the ultimate value of the cumulative-infection curve, including at the age-group level.
	This agreement supports the theory and enables direct prediction of the final state under vaccination, providing a foundation for optimizing vaccine allocation.
	
	With this predictive capability, we can employ optimization algorithm to identify the optimal allocation that minimizes the final epidemic burden (Fig.~\ref{fig:strategy}d).
	Because we can predict, at any time, the final-state cumulative infections for each age group, 
	we are able to compute not only the total cumulative infections for the entire population but also total deaths and YLL, 
	using age-stratified IFRs ($\boldsymbol{\epsilon}$) and remaining life expectancy ($\boldsymbol{\phi}$).
	This enables us to define, at any time, a cost function that maps a candidate vaccine allocation to a final epidemic burden (e.g., total cumulative infections, deaths, or YLL), 
	and to minimize this objective under practical constraints (e.g., limited vaccine supply) using an optimization algorithm such as Sequential Least Squares Programming (SLSQP).
	For instance, our optimized allocation is compared with five empirical strategies, demonstrating superior performance.
	
	The foregoing procedure addresses a one-shot allocation at a single time, whereas real-world vaccination is continuous and policies specify daily allocations across age groups. 
	To operate continually, our dynamic vaccine prioritization is then generated by implementing this optimization procedure within a Receding Horizon Control (RHC, also known as Model Predictive Control, MPC) framework (Fig.~\ref{fig:strategy}e).
	In details, at each vaccination decision time (daily in this study), subject to vaccine-supply constraints, 
	we utilize our final-state optimization method to determine and administer the age-stratified vaccine allocation that minimizes the real-time predicted final epidemic burden at that time, then repeat at the next vaccination time.
	For illustration, the example in Fig.~\ref{fig:strategy}e shows the time evolution of cumulative infections with three successive vaccination rounds (red curve). 
	Yellow, blue, and red arrows mark the final-state predictions made at the first, second, and third vaccination times, respectively. 
	Each prediction is based solely on information available up to that time and matches the realized final size after the corresponding number of rounds 
	(i.e., the prediction at the first vaccination time matches the final state after only one round of vaccination; at the second, after only two rounds; and so forth).
	This demonstrates that, in our dynamic vaccine prioritization, the optimization at each decision point only minimizes the final epidemic burden predicted at that time, which reflects the core principle of RHC.
	In this paper, we refer to this adaptive, prediction-based vaccination strategy as Final-state Dynamic Vaccine Prioritization (FS-DVP).

	\subsection*{Robustness from Dynamic Adaptation}
\begin{figure}[htbp]
		\centering
		\includegraphics[width=\linewidth]{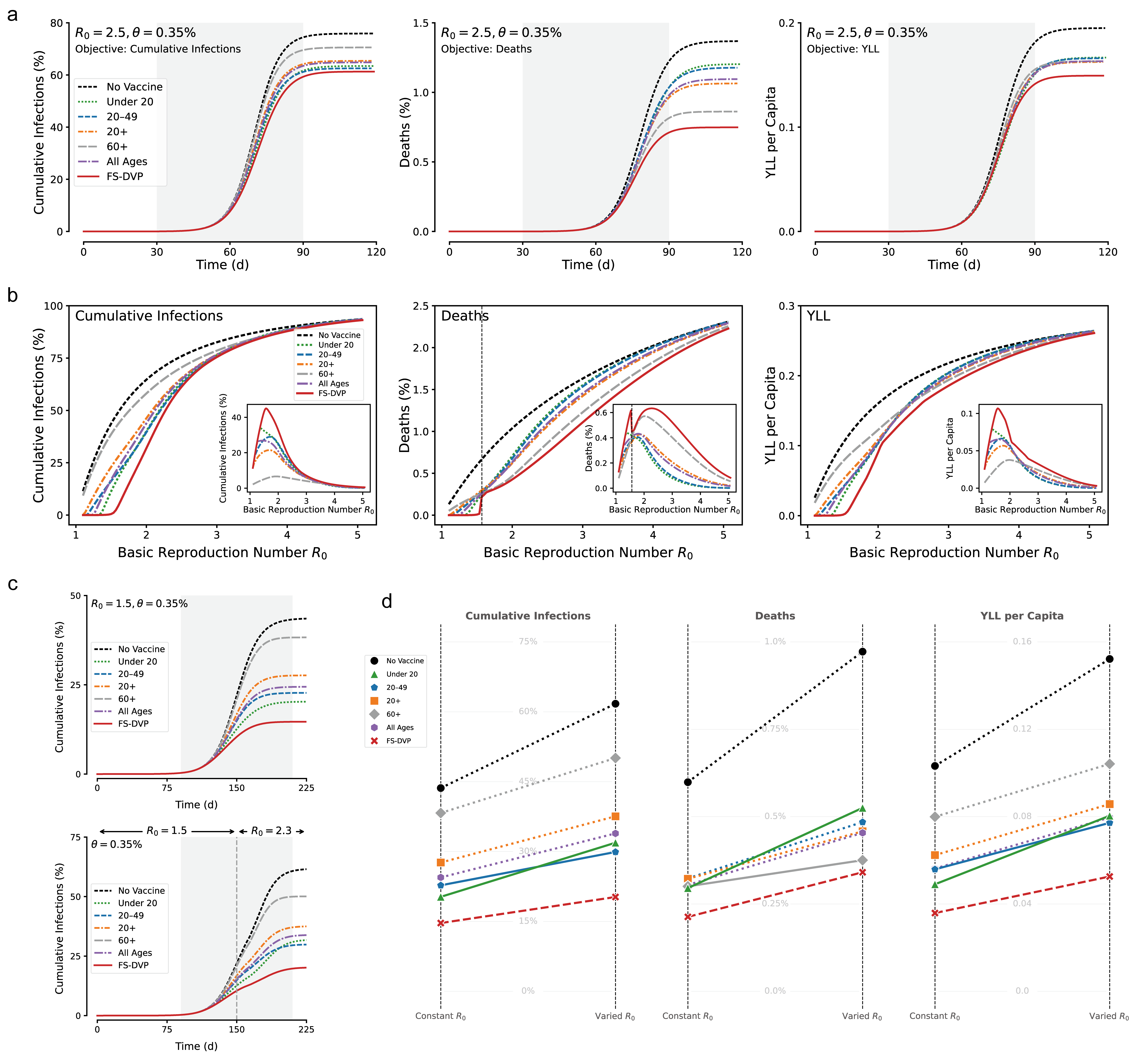}
		\caption{
			\textbf{a} Epidemic trajectories (cumulative infections, deaths, YLL) under non-Markovian final-state dynamic vaccine prioritization (FS-DVP, red solid curve) versus static prioritization strategies targeting specific age groups, for $R_0 = 2.5$, $\theta = 0.35\%$, and a 60-day vaccination campaign. 
			The gray shaded region indicates the vaccination period.
			\textbf{b} Final-state cumulative infections, deaths, and YLL as a function of $R_0$, comparing FS-DVP to five empirical static strategies; insets show the difference from no-vaccination baseline. Vertical dotted line in middle panel indicates where the performance ranking of static strategies changes substantially for reducing deaths.
			\textbf{c} Time evolution of cumulative infections for static versus dynamic allocation when $R_0$ is constant (upper: $R_0 = 1.5$) or changes mid-outbreak (lower: $R_0$ shifts from 1.5 to 2.3 at day 150). The gray shaded region indicates the vaccination period.
			\textbf{d} Final-state outcomes for each control objective (cumulative infections, deaths, YLL) under both constant and varied $R_0$ scenarios. FS-DVP (red dashed) consistently achieves the lowest epidemic burden, while the optimal static strategy varies with scenario and objective.}
		\label{fig:dy_robustness}
	\end{figure}

	To systematically assess the robustness of FS-DVP arising from dynamic adaptation, we evaluate its performance against empirical static allocation strategies.
	As shown in Fig.~\ref{fig:dy_robustness}a, for a baseline scenario with $R_0 = 2.5$ and daily vaccination rate $\theta = 0.35\%$ sustained over $60$ days, 
	our FS-DVP framework dynamically generates optimal vaccine allocations tailored to the chosen control objective, cumulative infections (left panel), deaths (middle panel), or YLL (right panel). 
	In every case, FS-DVP achieves superior epidemic control compared to static strategies that prioritize fixed age groups (under 20, 20–49, 20+, 60+, or all ages), as proposed in Ref.~\cite{bubar2021}.
	
	Fig.~\ref{fig:dy_robustness}b extends this analysis across a wide range of $R_0$ values. 
	FS-DVP consistently outperforms all static strategies across the full spectrum of $R_0$, 
	highlighting its adaptability and resilience to changing transmission scenarios.
	Notably, among the five static strategies, performance shifts with the underlying transmission rate: 
	prioritizing children and adolescents (under 20) yields the best outcomes at low $R_0$, 
	whereas targeting adults (20–49) or older adults (60+) becomes preferable as $R_0$ increases, depending on the objective.
	This sensitivity implies that a static allocation performing well under one transmission scenario may perform poorly under another, 
	underscoring the inherent limitations of fixed strategies in dynamic epidemic settings.
	
	We also evaluated the robustness of FS-DVP under time-varying transmission conditions (Fig.~\ref{fig:dy_robustness}c–d). 
	In Fig.~\ref{fig:dy_robustness}c, we compare the performance of static strategies and FS-DVP in reducing cumulative infections, 
	both when $R_0$ remains constant throughout the epidemic (upper panel) and when $R_0$ increases partway through the outbreak (lower panel; potentially due to viral mutation or changes of human behavior). 
	In both cases, FS-DVP consistently achieves the lowest cumulative infections. 
	For static strategies, the optimal choice shifts depending on the scenario: 
	vaccinating those under 20 is best when $R_0$ is constant, but when $R_0$ increases, 
	prioritizing adults aged 20–49 becomes more effective, illustrating the lack of flexibility in static approaches.
	
	Fig.~\ref{fig:dy_robustness}d further summarizes the final epidemic burdens (cumulative infections, deaths, and YLL) across both constant and changing $R_0$ scenarios for all strategies. 
	FS-DVP remains optimal across all objectives, while the best static strategy varies: 
	for cumulative infections, the optimal static target shifts from under 20 to 20–49; 
	for deaths, from under 20 to 60+; and for YLL, from under 20 to 20–49. 
	These results highlight that static strategies are highly sensitive to changing epidemic conditions and may fail to remain effective as the context shifts, 
	whereas FS-DVP adapts in real time and consistently delivers the best or near-best outcomes.
	(We also demonstrate the robustness of our FS-DVP across different vaccination campaign durations, daily vaccine supplies, and vaccine efficacies; see Supplementary Note 4 for details.)
	
	\subsection*{Robustness from Final-state Optimization}
	
	\begin{figure}[htbp]
		\centering
		\includegraphics[width=\linewidth]{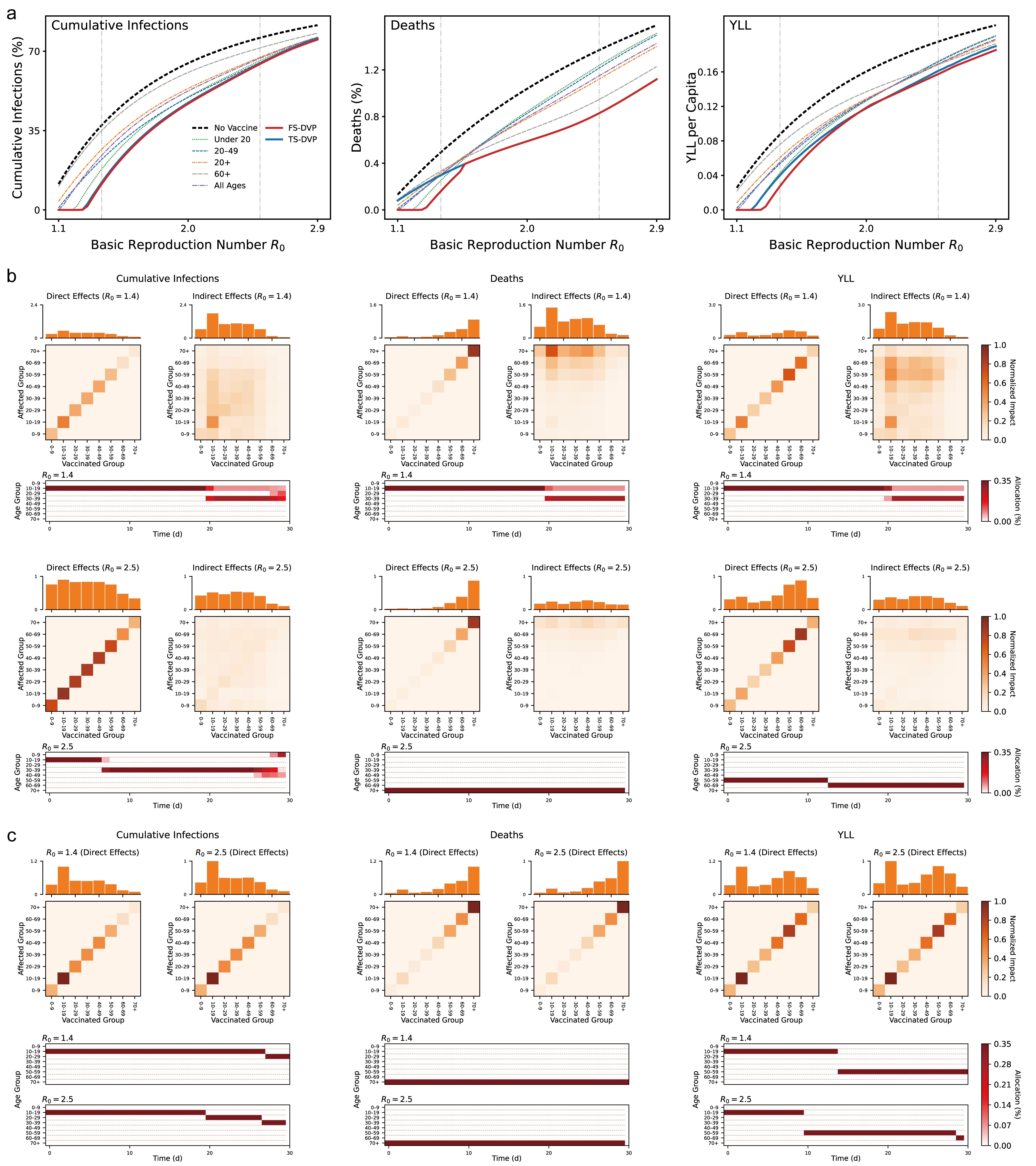}
		\caption{
			\textbf{a} Comparison of final epidemic burdens, i.e., cumulative infections, deaths, and years of life lost (YLL), under a non‐Markovian model across a range of $R_0$ values (from $1.1$ to $2.9$). 
			The primary comparison is between two dynamic strategies: Final‐State Dynamic Vaccine Prioritization (FS‐DVP, red solid) and Transient‐State Dynamic Vaccine Prioritization (TS‐DVP, blue solid).
			Their performance is benchmarked against static strategies and a no‐vaccination baseline. 
			Vertical dashed lines indicate $R_0=1.4$ and $2.5$, which are the values used for analysis in (\textbf{b})--(\textbf{c}). 
			\textbf{b} The direct and indirect effects detected by FS-DVP and the resulting optimal, age-stratified dynamic vaccine allocations.
			Panels are arranged by objective (columns: minimizing cumulative infections, deaths, YLL) and transmission level (rows: $R_0 = 1.4$, $2.5$). 
			In each panel, the top two heatmaps show the direct and indirect effects detected by FS-DVP (entries normalized by the maximum value of the total-effects matrix, i.e., the combined direct and indirect effects, with embedded marginal bars indicating column sums), 
			and the temporal heatmap below shows the age-stratified, dynamic allocations produced by FS-DVP.
			\textbf{c} The direct effects detected by TS-DVP and the resulting optimal, age-stratified dynamic vaccine allocations.
			Panels are arranged by objective (columns: minimizing cumulative infections, deaths, YLL). 
			Because TS-DVP lacks access to indirect effects, each panel displays two heatmaps of direct effects detected by TS-DVP at $R_0 = 1.4$ and $R_0 = 2.5$ 
			(each matrix max-normalized to $1$ with embedded column-sum bars), and the two heatmap below show the age-stratified, 
			dynamic vaccine allocations produced by TS-DVP at $R_0 = 1.4$ and $2.5$.}
		\label{fig:nf_robustness}
	\end{figure}

	To facilitate subsequent analysis, we distinguish direct/indirect effects from direct/indirect protection. 
	Effects describe the causal impact after vaccinating individuals, whereas protection refers to prioritization strategies designed to meet a specified objective.
	Any vaccination strategy can generate both direct and indirect effects: direct effects are the impact of vaccinating an individual on that same individual, 
	whereas indirect effects are the impact of vaccinating one person on unvaccinated others (e.g., by reducing transmission)~\cite{halloran1997study,fine2011herd}.
	A typical example of indirect effects is herd immunity: when a sufficient proportion of individuals acquire immunity, disease transmission is impeded and may be substantially reduced or even halted~\cite{fine2011herd}.
	By contrast, protection concerns how vaccination strategies are designed to meet a specified objective:
	direct protection strategies prioritize groups primarily based on direct effects and therefore focus on high-importance groups, 
	while indirect protection strategies prioritize groups mainly based on indirect effects, therefore focusing on high-contact groups.
	For direct protection, the ``importance'' of each age group is defined by the control objective: 
	to minimize cumulative infections, prioritize younger groups with high transmission potential; 
	to minimize deaths, prioritize older adults with higher infection-fatality risk (the conventional high-risk groups); 
	and to minimize YLL, prioritize those who contribute most to YLL, typically adolescents/young adults and middle-aged adults. 
	For indirect protection, the emphasis is on suppressing transmission by achieving high coverage in high-contact groups. 
	Notably, when the objective is to minimize cumulative infections or YLL, high-contact and high-importance groups often coincide or substantially overlap, 
	so vaccinating younger cohorts can simultaneously serve as both a direct- and an indirect-protection strategy.
	
	Previous work on dynamic vaccine allocation has largely focused on transient-state optimization, minimizing short-term transmission while ignoring long-term epidemic outcomes, as in Ref.~\cite{han2021time}; 
	we refer to this class as transient-state dynamic vaccine prioritization (TS-DVP). 
	To systematically assess the robustness provided by final-state optimization, we compare FS-DVP with TS-DVP across a range of $R_0$,
	and evaluate their performance in reducing cumulative infections, deaths, and YLL, alongside empirical static strategies for reference. 
	In our study, TS-DVP is implemented by setting its optimization objective to minimize the model-derived increase in epidemic burden at the time when vaccine-induced immunity manifests (see Methods for details).

	Furthermore, we analyze how the direct and/or indirect effects of vaccination influence the dynamic vaccine schedules generated by these two strategies.
	All vaccine allocations, including those under FS-DVP and TS-DVP, produce both direct and indirect effects: direct effects arise immediately, whereas indirect effects require more time to manifest.
	The key difference between the two strategies is whether these effects of candidate vaccine allocations are considered during optimization: 
	FS-DVP evaluates candidate vaccine allocations over a longer horizon and can therefore account for both direct and indirect effects during optimization, 
	whereas TS-DVP accounts solely for immediate outcomes and thus captures only the direct effects of candidate vaccine allocations, with indirect effects typically overlooked.
	Therefore, as we will show below, FS-DVP explicitly trades off direct and indirect effects under varying transmission conditions, enabling adaptive adjustment between direct and indirect protection.
	In contrast, TS-DVP is constrained by its short-term objective and therefore only prioritizes direct protection.
	
%

	As shown in Fig.~\ref{fig:nf_robustness}a, we observe the following pattern across objectives. For minimizing cumulative infections, FS-DVP only marginally outperforms TS-DVP. 
	For minimizing deaths, FS-DVP provides a substantial advantage when $R_0$ is low; in this regime, TS-DVP performs poorly, even underperforming some static strategies.
	For minimizing YLL, FS-DVP consistently surpasses TS-DVP. And Figs.~\ref{fig:nf_robustness}b--c reveal the detailed mechanism underlying this pattern.
	Figs.~\ref{fig:nf_robustness}b--c show, for $R_0=1.4$ and $R_0=2.5$, the direct and indirect effects detectable under FS-DVP and TS-DVP (TS-DVP has access only to direct effects), 
	along with the corresponding dynamic vaccine allocation schedules produced by each strategy, 
	when optimizing distinct objectives (see Methods for the calculation of direct and indirect effects).
	
	To interpret the patterns in Fig.~\ref{fig:nf_robustness}, we summarize how FS-DVP behave across values of $R_0$ when minimizing different objectives. 
	When $R_0$ is small, outbreaks are easier to control and blocking transmission is highly effective, 
	so vaccinating high-contact younger groups provides indirect protection that better shields high-importance populations, highlighting the increasing relative contribution of indirect vaccination effects (Fig.~\ref{fig:nf_robustness}b).
	Thus, in this low-$R_0$ regime, one principle follows: indirect effects dominate, and policy emphasizes indirect protection by prioritizing vaccination of high-contact age groups.
	As $R_0$ increases, control becomes harder and transmission is more difficult to interrupt, which suppresses total indirect effects; 
	moreover, the delay to vaccine-induced immunity means high-contact individuals are more likely to be infected before protection develops, which reduces their relative direct benefit.
	Consequently, in the high-$R_0$ regime, these two phenomena give rise to two principles: 
	(i) direct effects dominate, and policy emphasizes direct protection by prioritizing vaccination of high-importance age groups;
	(ii) as much as possible, policy avoids vaccinating those who are high-contact (Fig.~\ref{fig:nf_robustness}b).
	Therefore, in high $R_0$ setting, dynamic vaccine allocation then depends on the objective:
	to minimize cumulative infections, because high-importance and high-contact groups largely overlap, allocation may start with the highest-importance age group (e.g., 10–19, also with high contacts) and quickly shift to groups with slightly lower importance rates (e.g., 30–39, with lower contacts);
	to minimize deaths, as high-importance and high-contact groups do not overlap at all, older adults with high infection-fatality rates (e.g., 70+) are prioritized;
	to minimize YLL, since high-contact groups form part of the broader high-importance category, middle-aged adults with high importance (e.g., 50–59 and 60–69) are prioritized, while vaccination of high-contact, high-importance youth (e.g., 10–19) is de-emphasized.

	By contrast, TS-DVP focuses on short-term reduction, and because indirect effects do not materialize quickly enough after vaccinating a group, it effectively relies only on direct effects (Fig.~\ref{fig:nf_robustness}c).
	Therefore, across all values of $R_0$, TS-DVP vaccinates high-importance groups to provide direct protection.
	
	These design differences yield the following patterns. 
	For minimizing cumulative infections, both FS-DVP and TS-DVP consistently prioritize youth across all $R_0$ values and achieve similar outcomes, with FS-DVP performing only slightly better.
	For minimizing deaths, FS-DVP favors indirect protection when $R_0$ is low and clearly outperforms TS-DVP; as $R_0$ increases, both strategies provide direct protection, resulting in comparable performance.
	For minimizing YLL, FS-DVP also favors indirect protection and markedly outperforms TS-DVP at low $R_0$;
	at higher $R_0$, although both policies tend to emphasize direct protection, FS-DVP evaluates direct effects from a long-term perspective and accounts for the relative decline in direct benefits among youth as $R_0$ rises, 
	enabling it to maintain superior performance over TS-DVP even when only direct effects are considered.
	In summary, FS-DVP incorporates both direct and indirect effects, while TS-DVP is effectively limited to direct effects alone. As a result, FS-DVP achieves superior overall control, favoring indirect protection when $R_0$ is low and shifting toward direct protection as $R_0$ increases.
	
	Note that although FS-DVP identifies the optimal allocation strategy at each vaccination decision point, each optimization does not take the future vaccination rounds into consideration. 
	Consequently, the allocation chosen at each decision time point may be only locally optimal over the entire dynamic vaccination process. Hence, NF-DVP is not universally optimal: 
	for example, in scenarios aiming to minimize YLL at intermediate transmission levels, when the vaccination campaign is prolonged, TS-DVP may perform slightly better (see Supplementary Note 5 for details).

	\subsection*{Mechanistic Insights of the Dynamic Optimal Vaccine Prioritization}
	
\begin{figure}[htbp]
		\centering
		\includegraphics[width=\linewidth]{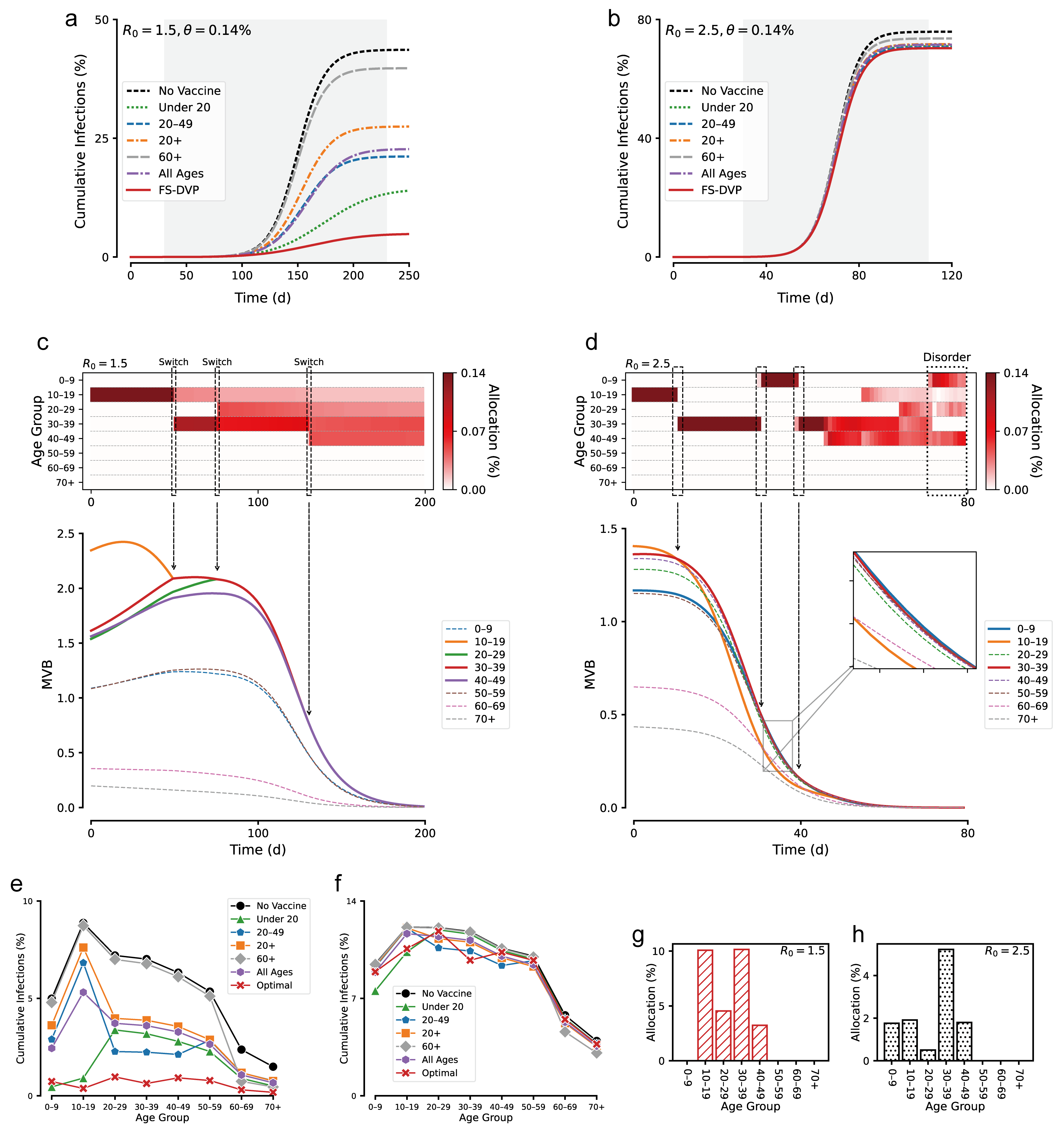}
		\caption{\textbf{a}--\textbf{b} Comparison of the time evolution curves for the cumulative infected fraction under different strategies, with the basic reproduction number $R_0$ set to $1.5$ (\textbf{a}) and $2.5$ (\textbf{b}). The daily vaccine supply $\theta$ is $0.14\%$ of the total population, and the gray area indicates the vaccination period of 200 days (\textbf{a}) and 80 days (\textbf{b}). \textbf{c}--\textbf{d} The upper panels display our dynamic optimal vaccine allocations over time, with three dashed rectangles marking the first three vaccine shift time points. The dotted area in \textbf{d} highlights the disorder during the vaccination period. The lower panels show the marginal vaccination benefit (MVB) $\xi^{\mathrm{c}}_l$ for each age group $l$ over time, and arrows indicating where the highest MVB curves, together with shifts in vaccine allocation, merge when $R_0$ equals $1.5$ with smaller value (\textbf{c}) or intersect when $R_0$ equals $2.5$ with larger value (\textbf{d}). Inset in the lower panel of \textbf(d) zooms in on the MVB curves when only vaccinating the age group 0–9 between the second and third shift time points, demonstrating the highest MVB value for 0–9 during that period. \textbf{e}--\textbf{f} Show the final‐state cumulative infected fraction of each age group relative to the total population under different strategies with $R_0$ set to $1.5$ (\textbf{e}) and $2.5$ (\textbf{f}). \textbf{g}--\textbf{h} illustrate the sum of the time‐varying optimal vaccine allocation for each age group.}
		\label{fig:mechanism}
	\end{figure}

	We have investigated the mechanism by which FS-DVP yields different allocation patterns under varying levels of $R_0$. 
	Additionally, even under a fixed $R_0$, the optimal vaccine allocation evolves dynamically as the epidemic progresses and coverage accumulates.
	To explore the mechanistic insights behind shifts in vaccine prioritization as the epidemic unfolds and vaccination progresses,
	we conduct simulations under two scenarios: $R_0 = 1.5$ and $R_0 = 2.5$, each with a daily vaccine supply $\theta$ equal to $0.14\%$ of the total population, as shown in Fig.~\ref{fig:mechanism}a--b. 
	
	The dynamic optimal strategy exhibits distinct patterns of partial or full allocation shifts between age groups. As shown in Fig.~\ref{fig:mechanism}c--d, for $R_0 = 1.5$, initial prioritization of the 10–19 age group transitions with partial shifts: at the 50-th day of vaccination, a portion of vaccines reallocates to 30–39 while maintaining coverage for 10–19; subsequent shifts to 20–29 and 40–49 occur at days 76 and 131, respectively. In the $R_0 = 2.5$ scenario, a more drastic pattern emerges: initial focus on 10–19 shifts entirely to 30–39 at day 11, followed by complete switches to 0–9 and later reintroduces 10–19, sometime later, alongside partial allocations to 40–49 and 20–29. Late-stage allocations become disordered as vaccine effectiveness diminishes in influencing transmission dynamics, because at this time there is no much optimization potential. 
	
	To elucidate the two different patterns, i.e., the partial versus full reallocation of vaccines, we define the marginal vaccination benefit (MVB) for each age group as the marginal reduction per unit dose in a chosen final-state objective, 
	such as cumulative infections, deaths, or YLL (denoted by the vectors $\boldsymbol{\xi}^{\mathrm{c}}$, $\boldsymbol{\xi}^{\mathrm{d}}$, and $\boldsymbol{\xi}^{\mathrm{y}}$, respectively). 
	The MVB is obtained by allocating an additional dose exclusively to a given age group and evaluating, in the small-dose limit, the resulting reduction in the predicted final burden (see Methods for details of the calculation).
	This metric quantifies the final-state epidemiological benefit of vaccinating each age group at a given time point, serving as a dynamic guide for optimal vaccine allocation decisions.
	Optimization algorithms tend to prioritize age groups with the highest MVB, which dynamically evolves based on both transmission dynamics and vaccination processes, 
	where intersections or merges of MVB curves trigger allocation switches, as shown in Figs.~\ref{fig:mechanism}c--d for the case of minimizing cumulative infections (results for deaths and YLL are provided in Supplementary Note 6).
	
	When $R_0$ is small (Fig.~\ref{fig:mechanism}c), vaccination and transmission exert comparable influences on MVB. As the epidemic evolves, the highest two competing MVB curves of groups 10–19 and 30–39 can merge, indicating their marginal containment utilities become similar. 
	If the optimizer fully prioritizes one group, the neglected group’s MVB rapidly surpasses that of the prioritized one (due to rising infection pressure), making the strategy not optimal. 
	To maximize utility, the algorithm partitions doses between both groups, balancing their competing priorities and resulting in a partial switch (e.g., simultaneous allocation to 10–19 and 30–39).
	
	At higher $R_0$ (Fig.~\ref{fig:mechanism}d), rapid transmission dominates, weakening vaccination’s relative impact, leading highest MVB curve to intersect with a competing group’s curve (e.g., 10–19 $\to$ 30–39; because rapid transmission dominates, prioritizing one group does not allow the neglected group’s MVB to surpass that of the prioritized one), the optimizer abruptly shifts all doses to the newly dominant group. Even when several groups have nearly the same MVB values, such as during the second and third allocation shifts shown in the inset of Fig.~\ref{fig:mechanism}d, the algorithm still selects the one with the slightly highest MVB (e.g., 0–9). As the epidemic progresses, infection saturates across groups, and all MVB values fall to near zero. At this point, vaccination has little effect on transmission, the MVB curves flatten and converge, and the strategy shifts to distributing doses more evenly across groups. 
	We also provide a mathematical analysis of the above MVB-related mechanism (see Supplementary Note 7 for details).
%
	
	Notably, while the optimal strategy minimizes overall infections, single age-group outcomes may not surpass those of empirical strategies. 
	As shown in Fig.~\ref{fig:mechanism}e–f, under $R_0 = 1.5$, infections in the 0–9 group remain higher than in ``prioritize under 20'' strategies. For $R_0 = 2.5$, only the 30–34 group achieves lower infections compared to all the empirical approaches. 
	This reflects the strategy's differential prioritization: aggregated allocations (Fig.~\ref{fig:mechanism}g–i) reveal concurrent targeting of 10–19 and 30–39 for $R_0 = 1.5$ versus exclusive focus on 30–39 for $R_0 = 2.5$, driven by distinct MVB dynamics under varying transmission intensities.
	
	Above all, the concept of MVB explains why and how vaccine allocation shifts between age groups, 
	thereby not only deepening our understanding of the mechanisms underlying dynamic prioritization but also offering practical guidance for adapting vaccination policies in real-world settings.
	Additionally, detecting switching points allows us to divide the vaccination horizon into intervals, 
	thereby coarse-graining dynamic vaccine prioritization into strategies that remain constant within each interval, 
	which reduces implementation complexity (see Supplementary Note 8 for details).

	\subsection*{Application: Dynamic Vaccine Prioritization in the COVID-19 Pandemic}
	
	The study of disease control strategies is fundamentally aimed at guiding real-world interventions. 
	In the early stages of an outbreak, this requires accurate estimation of key epidemiological parameters, 
	followed by simulation of the outbreak trajectory using these estimates to inform timely and effective response plans (see Methods for details).
	Using COVID-19 as a case study, we examine the impact of vaccination strategies across nine countries, 
	i.e., Ireland, Japan, the United Kingdom, Singapore, France, Italy, Germany, the United States, and Spain, 
	by comparing epidemiological outcomes with and without vaccination, assuming a daily rollout rate of $0.35\%$ sustained over a 120-day period.
	(see Supplementary Notes 9--12 for the detailed epidemiological and demographic parameters of these countries). All results are reported with corresponding 95\% confidence intervals (CI).
	Note that, in reality, once vaccination is introduced, the corresponding no-vaccination trajectory cannot be observed; here it is provided as a counterfactual generated by our model for illustrative purposes, 
	and the results are not intended as a direct comparison with real-world vaccination practices.
	
	\begin{figure}[htbp]
		\centering
		\includegraphics[width=\linewidth]{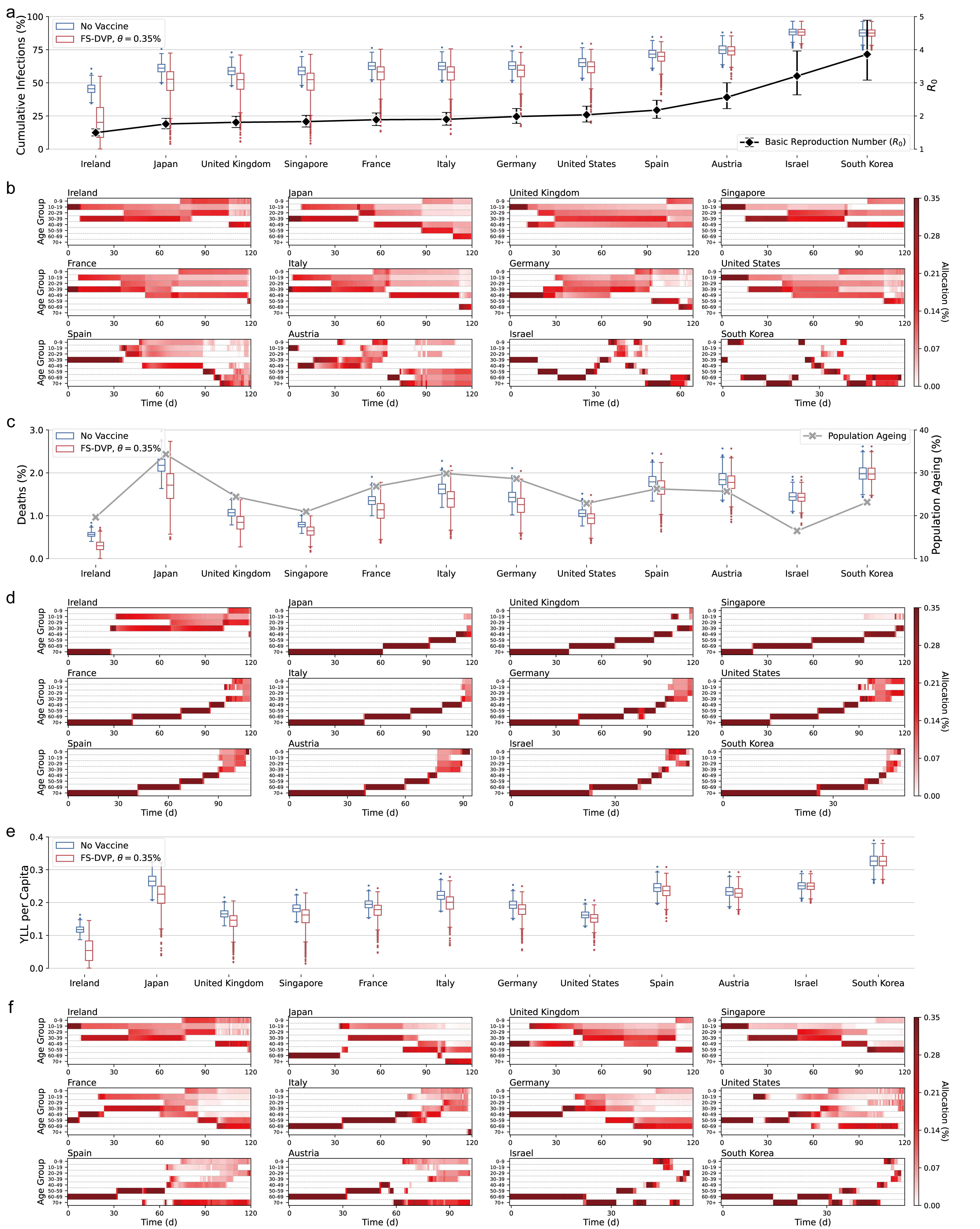}
		\caption{
			\textbf{a} Cumulative infection fractions (95\% CI) across countries under no vaccination and the time-varying optimal strategy. Countries are ordered by ascending $R_0$, shown by the black curve with diamond markers (95\% CI).
			\textbf{b} Time-varying optimal vaccine allocations by age group for minimizing cumulative infections across countries.
			\textbf{c} Death fractions (95\% CI) under no vaccination and the time-varying strategy. The gray curve with $\times$ markers shows the ageing index (95\% CI) for each country.
			\textbf{d} Time-varying optimal vaccine allocations by age group for minimizing deaths across countries.
			\textbf{e} Years of life lost (YLL) per capita (95\% CI) under no vaccination and the time-varying strategy.
			\textbf{f} Time-varying vaccine allocations by age group for minimizing YLL across countries.
			All results shown in panels (\textbf{a}, \textbf{c}, and \textbf{e}) are calculated from 1,000 independent simulation runs. For each simulation, the shape and scale parameters of the generation time distribution, as well as the basic reproduction number $R_0$, are randomly sampled for their distribution. In contrast, the time-varying optimal vaccine allocation results presented in panels (\textbf{b}, \textbf{d}, and \textbf{f}) are based on a single simulation performed with the mean values of these parameters. Other parameters are fixed as follows: daily vaccine supply $\theta = 0.35\%$, vaccination duration $T_{\mathrm{vac}} = 120$ days, and vaccine efficacy $\eta = 95\%$.}
		\label{fig:countries}
	\end{figure}
	
	Fig.~\ref{fig:countries} presents a comprehensive cross-country comparison of epidemic outcomes and optimal vaccine allocation strategies under different optimization objectives. 
	In Fig.~\ref{fig:countries}a, final cumulative infections increase with the $R_0$ baseline, as countries are ordered from left to right by ascending $R_0$. 
	As $R_0$ increases, the effectiveness of vaccination in suppressing cumulative infections diminishes, with the largest reductions observed in countries with lower $R_0$, 
	such as Ireland and Japan, while little to no reduction is seen in countries with high $R_0$, such as Israel and South Korea. 
	Furthermore, Figs.~\ref{fig:countries}c and e show that both deaths and YLL per capita are significantly reduced under the optimal vaccination strategy, particularly in countries with high population ageing or low $R_0$ 
	(population ageing is typically quantified by the share of the population aged 60 years or older).
	We further observe that CI is generally wider under the vaccination strategy than in the no-vaccination scenario. This is partly because, when $R_0$ is low, the optimal strategy can be highly effective in suppressing the epidemic, substantially lowering the lower bound of the CI. However, with higher $R_0$, the epidemic remains difficult to control, so the upper bound does not decrease as much, resulting in a broader overall CI. 
	
	
	Figs.\ref{fig:countries}b, d, and f illustrate the time-varying allocation of vaccines by age group under different optimization objectives. 
	Specifically, Fig.~\ref{fig:countries}b shows that, in most countries, minimizing cumulative infections is achieved by initially prioritizing vaccines for age groups 30–39, 10–19, and sometimes 40–49, 
	before partially reallocating to other age groups as the campaign progresses. 
	In countries with the highest $R_0$ values, such as Austria, Israel, and South Korea, the optimal strategy features a complete shift in vaccine allocation to other groups at a certain point in time.
	
	When minimizing deaths (Fig.\ref{fig:countries}d), most countries begin by vaccinating the 70+ age group, then fully reallocating vaccines to other groups; 
	Ireland stands out as the only country where the allocation shifts from the elderly to young groups, because of its low $R_0$. 
	For minimizing YLL (Fig.\ref{fig:countries}f), allocation patterns are more complex: in most countries, the initial focus is on the 40–49, 50–59, or 60–69 age groups, with subsequent redistribution to others, and no country vaccinates the 70+ group first. 
	Only in Ireland and Singapore does the optimal strategy start with the 10–19 age group, followed by partial allocation to those aged 30–39.
	
	Overall, these results demonstrate that both the effectiveness and the optimal allocation pattern of vaccination are strongly influenced by each country’s epidemiological and demographic characteristics, 
	as well as the chosen public health objective.

	\section*{Discussion}
	
	

	Designing vaccine policies that account for both direct and indirect effects in realistic, memory-dependent outbreaks requires long-horizon reasoning, not just short-term incidence reduction. 
	We introduce a dynamic vaccine-prioritization framework that minimizes final-state burden within a non-Markovian model and show consistent gains over static planning and strategies optimized only for short-term objectives, most notably for deaths and YLL. 
	A clear regime shift emerges: at lower transmission the policy favors indirect protection by vaccinating high-contact groups to suppress spread, 
	whereas at higher transmission it shifts toward direct protection of high-importance groups for their own benefit (e.g., older adults for deaths, middle-aged adults for YLL). 
	Meanwhile, we also introduce a diagnostic indicator, MVB, which tracks allocation switches over time and links reallocation decisions to both the epidemic phase and prevailing coverage levels.
	
	Our performance gains arise from a final-state predictor, which maps the non-Markovian system at any decision time to a final-state-equivalent Markovian surrogate, 
	combined with a lookahead sliding window that enables rapid, immediate evaluation of the long-term effectiveness of vaccination without forward simulation (Fig.~\ref{fig:strategy}a--c).
	Based on this predictor, our dynamic strategy continuously allocates vaccine doses to minimize the final epidemic burden, explicitly balancing indirect transmission blocking with the direct protection of high-importance groups.
	Prior studies have also demonstrated such an equivalence, but their results typically construct a Markovian counterpart only at the outbreak’s onset, rather than at arbitrary time points~\cite{feng2019equivalence,starnini2017equivalence,feng2023validity}.
	By introducing the residual effective infection rate, 
	we extend this equivalence to any intermediate time, which permits on-the-fly final-state prediction under vaccination and removes the need for costly forward simulations after each allocation decision.
	This predictor integrates naturally into a receding-horizon optimizer, allowing allocations to explicitly balance indirect transmission blocking with direct protection.
	
	While our non-Markovian formulation improves realism by capturing memory in transmission and protection, 
	it also admits state aggregation: when state progression of the dynamics is one-way (a directed acyclic state graph with absorbing classes), 
	the multiple stages of diverse epidemic models can be consolidated into a single state representation, 
	which enables our framework to encompass a wide range of model structures and thereby ensures that the proposed methods and conclusions remain broadly applicable.
	By contrast, in models with recurrent transitions (SIS, SIRS, or waning immunity with reinfection), the process approaches an endemic equilibrium rather than an absorbing final state~\cite{feng2019equivalence,starnini2017equivalence,van2013non,shaman2012forecasting,hethcote1989three}. 
	Under a finite vaccination window with protection that ultimately wanes, vaccination mainly reshapes the transient path (e.g., lowers peaks, delays convergence) but leaves the long-run equilibrium essentially unchanged; 
	only non-waning protection (as in our SIRP with absorbing $\textbf{P}$; Fig.1b–d) or sustained vaccination strong enough to keep the effective reproduction number below one would shift the long-run state~\cite{feng2019equivalence,starnini2017equivalence}. 
	Extending our approach to such settings is an important direction.
	
	While state aggregation gives our model strong adaptability to epidemic dynamics, the framework is also readily extensible along three aspects: population stratification, intervention types, and policy objectives.
	(i) Although we stratify by age in the main specification, the same machinery naturally extends to geography, occupation, or behavioral layers. 
	These schemes primarily reshape the contact structure, and hence the contact matrix, without compromising the model’s ability to predict the epidemic final state.
	(ii) Beyond vaccination, interventions such as lockdowns (which modify contacts) and quarantine (which alters infectiousness profiles) can be incorporated by adjusting the corresponding variables and then invoking the same final-state predictor.
	(iii) Real-world goals often go beyond minimizing cumulative infections, deaths, or YLL, and may include ICU occupancy, economic impact, and social equity.
	For example, the fairness of an intervention across different population groups is sometimes a critical factor, where it can be encoded via a fairness metric, either as a penalty term in the objective or as a hard constraint to ensure that minimum fairness thresholds are met.
	
	A key challenge in implementing dynamic vaccine prioritization strategies lies in their inherent flexibility: allocation recommendations shift continuously over time as epidemic conditions evolve. 
	While this dynamic nature enhances epidemiological effectiveness, it also introduces substantial logistical and managerial burdens in real-world settings.
	A practical way to operationalize dynamic vaccine prioritization is to let individuals register through an online system, and then assign vaccination times according to the dynamic allocation plan. 
	In this setup, the system does not simply record bookings but actively schedules registrants into time slots such that the distribution in each period aligns as closely as possible with the model’s evolving priorities. 
	This design reduces the need for disruptive on-site adjustments while ensuring that dynamic allocations are respected in aggregate.
	
	Our dynamic vaccine allocation framework can also inform the design of policies implemented in discrete intervals (e.g., weekly or monthly), 
	where allocations remain fixed within each interval to simplify scheduling and reduce operational complexity, as seen in schemes such as those proposed by Buckner et al.~\cite{buckner2021}. 
	This setting introduces two key challenges: how to determine the timing of allocation shifts, and how to choose the optimal allocation within each fixed period. 
	To address these challenges, we first compute the fully dynamic, time-continuous optimal allocation using our framework. 
	We then extract switching points from the dynamic allocation sequence to divide the rollout period into discrete phases, 
	and warm-start each phase-specific optimization using the corresponding portion of the dynamic solution,
	which could reduce convergence time while preserving the structure of the original dynamic strategy (see Supplementary Note 6 for detailed results).
	
	While many studies focus primarily on the direct effects of vaccination during an ongoing outbreak, some have incorporated indirect effects to a certain extent. 
	For example, Buckner et al. investigated optimal vaccine allocation across demographic groups over a six-month horizon~\cite{buckner2021}. 
	Within such a period, disease transmission may or may not reach the final state, yet indirect effects can still play a significant role. 
	However, their evaluation relied on forward simulations to assess vaccination outcomes, which are computationally intensive. 
	Moreover, because the system may not reach a true final state within the simulation window, the influence of indirect effects may not be fully captured.
	Although considering long-term impacts is crucial, some studies, constrained by computational complexity, adopt short-term objectives. 
	This does not mean that short-term goals are without value. 
	In certain scenarios, such as when healthcare resources are limited or ICU capacity is under strain, there is an urgent need to reduce the immediate burden on the healthcare system. 
	In such cases, short-term objectives are not only meaningful but necessary for effective policy design.
	
	In our study, FS-DVP does not always outperform TS-DVP: when $R_0$ is at a moderate level, near the threshold where the balance between indirect and direct protection shifts, 
	and the vaccination window is relatively long, TS-DVP can slightly outperform FS-DVP in minimizing YLL (see details in Supplementary Note 5). 
	Even in such cases, the performance difference is marginal, and the parameter regime in which TS-DVP is favored is very narrow.
	The reason for this phenomenon lies in the local nature of FS-DVP’s optimization. 
	Although FS-DVP identifies the optimal allocation strategy at each vaccination decision point, 
	it does not account for future vaccination rounds. 
	As a result, the vaccine allocation determined at each step may be locally optimal for the entire dynamic vaccination process.
	This limitation becomes particularly pronounced near the transition point between indirect and direct protection,
	where the algorithm may struggle to effectively balance the trade-off between immediate direct benefits and longer-term indirect effects for the full course of vaccination program.
	
	In our modeling framework, we assume that vaccine-induced protection becomes effective after a fixed delay, rather than modeling the onset of protection as a time-distributed process.
	This simplification enables the use of lookahead sliding window, allowing the model to directly compute how many individuals will become protected without relying on forward simulation. 
	In contrast, a time-distributed formulation would offer more detail but render the lookahead method infeasible, requiring forward simulation to estimate vaccine effectiveness and significantly increasing computational complexity.
	Developing a final-state predictor for models with time-distributed immunity remains an open methodological challenge.
	
	In sum, by making long-term effects computable within a non-Markovian setting and clarifying when and why priorities should switch, 
	our approach lays a foundation for adaptive vaccine policy; the extensions above aim to broaden its applicability, robustness, and operational readiness.

	\section*{Method}
	\subsection{Time distributions and the corresponding survival and hazard functions}
	In our non-Markovian framework, the memory-dependent nature of infection and removal processes is characterized by hazard functions $\omega(\tau)$.
	We use subscripts ``inf'' and ``rem'' to distinguish the two processes: for example, $\omega_{\mathrm{inf}}(\tau)$ and $\omega_{\mathrm{rem}}(\tau)$ represent the infection and removal hazard functions, respectively.
	These hazard functions specify the instantaneous rate at which the corresponding event occurs at each infection age, and thus capture essential memory effects of the disease progression.
	Besides the hazard function, the non-Markovian characteristics of the infection and removal processes can also be described using the survival function $\Psi(\tau)$ and the event time distribution $\psi(\tau)$.
	Specifically, the survival function is calculated as $\Psi(\tau) = \exp\left(-\int_{0}^{\tau}{\omega(\tau')d\tau'}\right)$, which gives the probability that the corresponding event has not occurred during the infection age interval $[0, \tau)$.
	The event time distribution is $\psi(\tau) = \omega(\tau)\exp\left(-\int_{0}^{\tau}{\omega(\tau')d\tau'}\right)$, representing the probability density that the event occurs at infection age $\tau$.
	These three quantities, $\omega(\tau)$, $\Psi(\tau)$, and $\psi(\tau)$, are fully interconvertible, and we also have the following relationships: $\psi(\tau) = - d\Psi(\tau)/d\tau$, $\omega(\tau) = - d\ln\left(\Psi(\tau)\right)/d\tau$, $\Psi(\tau) = 1 - \int_{0}^{\tau}{\psi(\tau')d\tau'}$, $\omega(\tau) = \psi(\tau) / \left(1 - \int_{0}^{\tau}{\psi(\tau')d\tau'}\right)$.
	
	In our numerical simulations, we frequently employ the Weibull distribution as the event time distribution to model infection and removal dynamics, owing to its flexibility in capturing both increasing and decreasing hazard rates. 
	The Weibull distribution is defined as $\psi(\tau) = \frac{\alpha}{\beta}(\frac{\tau}{\beta})^{\alpha - 1}\exp\left(-(\frac{\tau}{\beta})^{\alpha}\right)$, where $\alpha$ and $\beta$ denote the shape and scale parameters, respectively.

	\subsection{Dynamical Equations}
	We define the time-dependent vectors
	$\boldsymbol{s}(t)$, $\boldsymbol{i}(t)$, $\boldsymbol{r}(t)$, $\boldsymbol{c}(t)$, and $\boldsymbol{p}(t)$
	to represent the fractions of susceptible, infected (prevalent), removed, cumulative infected, and protected individuals at time $t$, respectively.
	The vector $\boldsymbol{\vartheta}_t$ denotes the vaccine allocation at time $t$.
	
	In the absence of vaccination at time $t$, the non-Markovian spread is governed by the deterministic integro-differential system:
	\begin{align}
		\frac{d\boldsymbol{c}(t)}{dt} = \boldsymbol{s}(t)\odot\boldsymbol{j}(t),
		\label{eq:c_t}
	\end{align}
	where
	\begin{align}
		\boldsymbol{j}(t) = k\boldsymbol{A}\operatorname{diag}(\boldsymbol{\rho})\int_{0}^{t} {\omega_{\mathrm{inf}}(t - t') \Psi_{\mathrm{rem}}(t - t') \, d\boldsymbol{c}(t')}.
		\label{eq:j_t}
	\end{align}
	Here $\boldsymbol{j}(t)$ is the per-step infection pressure (force of infection), $k>0$ is a scaling constant, $\boldsymbol{A}$ is the contact matrix (rows: receivers, columns: sources), and $\boldsymbol{\rho}$ is the age distribution.
	Unless stated otherwise, $\odot$ and $\oslash$ denote element-wise multiplication and division, and matrix–vector products are written by juxtaposition.
	The dynamics of the susceptible, infected, and removed fractions are described as follows: $d\boldsymbol{s}(t)/dt = -\boldsymbol{s}(t)\odot\boldsymbol{j}(t)$, 
	$\boldsymbol{i}(t) = \int_{0}^{t} {\Psi_{\mathrm{rem}}(t-t^{\prime})\,d\boldsymbol{c}(t')}$, and $\boldsymbol{r}(t) = \int_{0}^{t} {\left(1-\Psi_{\mathrm{rem}}(t-t^{\prime})\right)\,d\boldsymbol{c}(t')}$.
	
	Upon vaccination at time $t$ with allocation $\boldsymbol{\vartheta}_t$, the susceptible population at time $t+\delta$ is updated according to:
	\begin{align}
		\boldsymbol{s}(t+\delta) &\leftarrow \boldsymbol{s}(t+\delta)-\eta\boldsymbol{\vartheta}_t\odot\boldsymbol{\sigma}(t),
		\label{eq:susceptible_reduction}
	\end{align}
	where the factor $\boldsymbol{\sigma}(t) = \exp\left(-\int_{t}^{t+\delta}{\boldsymbol{j}(t')dt'}\right)$ accounts for probability that infection fails to occur during the lag $(t,\,t+\delta]$.
	The corresponding reduction in the susceptible compartment is transferred into the protected class, yielding
	$\boldsymbol{p}(t+\delta) \leftarrow \boldsymbol{p}(t+\delta)+\eta\boldsymbol{\vartheta}_t\odot\boldsymbol{\sigma}(t)$.
	
	\subsection{Final-state Prediction without Vaccination}
	As illustrated in Fig.~\ref{fig:strategy}a, consider a non-Markovian outbreak observed at time $t$.
	If an individual was infected at calendar time $t'$, then its current infection age is $a:=t-t'\ge 0$.
	Since the infection and removal hazards depend on infection age, $\omega_{\mathrm{inf}}(\tau)$ and $\omega_{\mathrm{rem}}(\tau)$, the segments on $0\le \tau<a$ no longer affect this individual’s future transmission; only the residual hazards for $\tau\ge a$ are relevant.
	We summarize the remaining transmissibility at infection age $a$ by the residual effective infection rate: 
	\begin{align}
		\lambda_{\dagger}(a) = \int_{a}^{\infty}{\omega_{\mathrm{inf}}(\tau)\exp\left(- \int_{a}^{\tau}{{\omega_{\mathrm{rem}}(\tau')d\tau'}}\right)d\tau}
		\label{eq:residual_effective_infection_rate}
	\end{align}
	Replacing the individual’s non-Markovian hazards by constants
	$\omega_{\mathrm{inf}}(\tau)\equiv\mu\,\lambda_{\dagger}(a)$ and
	$\omega_{\mathrm{rem}}(\tau)\equiv\mu$ (with arbitrary $\mu>0$) yields an equivalent
	Markovian surrogate whose final state coincides with that of the original non-Markovian process (time rescaling does not alter the final size).
	Using the Kermack–McKendrick final-size relation which solve the Markovian final state, the non-Markovian final state from time $t$ satisfies the transcendental equation:
	\begin{align}
	\tilde{\boldsymbol{c}}= \boldsymbol{1} - \boldsymbol{s}(t)\odot\boldsymbol{\zeta}(t), 
	\label{eq:non_vac_final_state}
	\end{align}
	where
	\begin{align}
		\boldsymbol{\zeta}(t) = \exp\left(-k\boldsymbol{A}\operatorname{diag}(\boldsymbol{\rho})\left(\lambda_{\mathrm{eff}}\tilde{\boldsymbol{c}} - \lambda_{\mathrm{eff}}\boldsymbol{r}(t) - \widehat{\boldsymbol{\lambda}}_{\mathrm{eff}}(t)\odot \boldsymbol{i}(t)\right)\right).
	\end{align}
	Here $\exp(\cdot)$ acts elementwise, and the function $\boldsymbol{\zeta}(t)$ represents the probability that the individual in each age group, who has never been infected until time $t$, 
	will remain uninfected throughout the remainder of the epidemic. Here, $\lambda_{\mathrm{eff}}$ represents the effective infection rate, calculated as $\lambda_{\mathrm{eff}}=\int_0^{+\infty}{\omega_{\textrm{inf}}(\tau)\Psi_{\textrm{rem}}(\tau)d\tau}$.
	The term $\widehat{\boldsymbol{\lambda}}_{\mathrm{eff}}(t)$ captures the per-case ``used-up'' infectiousness of prevalent infections at time $t$, and satisfies
	$\widehat{\boldsymbol{\lambda}}_{\mathrm{eff}}(t)=\lambda_{\mathrm{eff}} - \int_{0}^{t}{\int_{t-t'}^{\infty}{\omega_{\mathrm{inf}}(\tau)\Psi_{\mathrm{rem}}(\tau)d\tau}d\boldsymbol{c}(t')} \oslash\boldsymbol{i}(t)$.
	The final cumulative infection fraction, death fraction in the total population and YLL per capita, denoted by $\tilde{\chi}_\mathrm{c}$, $\tilde{\chi}_\mathrm{d}$ and $\tilde{\chi}_\mathrm{y}$, are calculated as
	$\tilde{\chi}_\mathrm{c} = \boldsymbol{\rho}^{\mathsf{T}}\tilde{\boldsymbol{c}}$, 
	$\tilde{\chi}_\mathrm{d} = (\boldsymbol{\rho}\odot\boldsymbol{\epsilon})^{\mathsf{T}}\tilde{\boldsymbol{c}}$, 
	and $\tilde{\chi}_\mathrm{y} =(\boldsymbol{\rho}\odot\boldsymbol{\epsilon}\odot \boldsymbol{\phi})^{\mathsf{T}}\tilde{\boldsymbol{c}}$,
	where $\boldsymbol{\epsilon}$ denotes the age-stratified IFRs, $\boldsymbol{\phi}$ is the remaining life expectancy each age group.
	
	\subsection{Final-state Prediction with Vaccination}
	According to Eq.~\eqref{eq:susceptible_reduction}, introducing vaccines at time $t$ necessitates simulating the process up to $t+\delta$ to determine the fraction of individuals who successfully acquire immunity. 
	To handle the $\delta$ delay between giving a vaccine at time $t$ and protection taking effect at $t + \delta$, we keep a rolling ``lookahead' buffer, a future cache that always covers the interval $(t, t + \delta]$. 
	By the time we are at $t$, the simulator has already precomputed how the epidemic would evolve up to $t + \delta$ and stored those states.
	When vaccination happens at $t$, we adjust the cached state at the maturity time $t + \delta$: we move into ``protected'' exactly those vaccinated individuals who, according to the precomputed trajectory, would still be susceptible at $t + \delta$.
	Each simulation step then (i) commits the next precomputed slice (for $t + \Delta t$) from the cache into the main time series, and (ii) extends the right edge of the buffer by one step so the window length stays at $\delta$. 
	This design lets us read off, immediately and consistently, how much protection a vaccination at $t$ will produce at $t + \delta$ without re-simulating the entire interval after every vaccination when we need to optimize the vaccine allocation
	(see Supplementary for the detailed pseudocode).
	
	At any given time $t$, we denote the cached values at $t+\delta$ by $\boldsymbol{s}^*(t)$, $\boldsymbol{i}^*(t)$, $\boldsymbol{r}^*(t)$, $\boldsymbol{c}^*(t)$, and $\boldsymbol{p}^*(t)$. 
	This setup allows for immediate calculation of the effective immunity fraction at $t+\delta$ due to vaccination at $t$, $\boldsymbol{\vartheta}^{\mathrm{eff}}_{t} = \eta\boldsymbol{\vartheta}_{t}\odot\boldsymbol{\sigma}(t)$, and
	$\boldsymbol{\sigma}(t)$ could also be calculated as $\boldsymbol{\sigma}(t) = \exp\left(-\int_{t}^{t+\delta}{\boldsymbol{j}(t')dt'}\right) = \left(\boldsymbol{s}((t+\delta)^-) - \boldsymbol{u}((t+\delta)^-)\right)\oslash\left(\boldsymbol{s}(t^-) - \boldsymbol{v}(t^-) - \boldsymbol{u}(t^-)\right)$,
	where $\boldsymbol{\mathit{v}}(t)$ and $\boldsymbol{\mathit{u}}(t)$ denote, respectively, the fractions of vaccinated individuals and those unprotected due to a failure to acquire immunity, and the superscript ``$-$'' indicates the left-hand limit, applied in cases where the function is discontinuous.
	
	Then the final state satisfies:
	\begin{align}
		\tilde{\boldsymbol{c}}=[1 - \boldsymbol{p}^*(t) - \boldsymbol{\vartheta}^{\mathrm{eff}}_{t}] - [\boldsymbol{s}^*(t)- \boldsymbol{\vartheta}^{\mathrm{eff}}_{t}]\odot\boldsymbol{\zeta}^*(t).
		\label{eq:vac_final_state}
	\end{align}
	where
	\begin{align}
		\boldsymbol{\zeta}^*(t) &= \exp\left(-k\boldsymbol{A}\operatorname{diag}(\boldsymbol{\rho})\left(\lambda_{\mathrm{eff}}\tilde{\boldsymbol{c}} - \lambda_{\mathrm{eff}}\boldsymbol{r}^*(t) - \widehat{\boldsymbol{\lambda}}^*_{\mathrm{eff}}(t)\odot \boldsymbol{i}^*(t)\right)\right).
		\label{eq:zeta_star}
	\end{align}
	Here, $\widehat{\boldsymbol{\lambda}}^{*}_{\mathrm{eff}}(t)=\lambda_{\mathrm{eff}} - \int_{0}^{t+\delta}{\int_{t+\delta-t'}^{\infty}{\omega_{\mathrm{inf}}(\tau)\Psi_{\mathrm{rem}}(\tau)d\tau}d\boldsymbol{c}(t')} \oslash\boldsymbol{i}^*(t)$.
	
	\subsection{Empirical Vaccination Methods}
	We compare our vaccination strategy against five commonly used empirical approaches that prioritize specific age groups, as outlined by Bubar et al.~\cite{bubar2021}. These strategies include: vaccinating individuals under 20 years of age (``under 20''), adults aged 20 to 49 years (``20–49''), all adults aged 20 years and above (``20+''), adults aged 60 years and above (``60+''), and all age groups with equal prioritization (``all ages''). In each approach, the designated priority group receives vaccinations proportionally until fully covered, after which the remaining doses are allocated to other groups in a similar proportional manner. For instance, the ``under 20'' strategy entails proportionally vaccinating individuals under 20 years first, followed by proportional vaccination of the remaining age groups once the initial group is fully vaccinated.
	
	\subsection{Calculation of Direct/Indirect Effects}
	We denote by $\bar{\boldsymbol{\varsigma}}^{\mathrm{nf}}_{t}$ the direct effects observable under the FS-DVP framework (rows: vaccinated group, columns: affected group).
	The key idea behind its construction is to quantify, for each vaccinated individual, the potential reduction in the objective that would result if the acquired immunity did not contribute to halting disease transmission but only protected the vaccinated individual.
	The proportion of susceptibles that decrease from the onset of vaccine effectiveness to the final state, relative to the number of susceptibles at the onset, corresponds to the fraction of susceptible individuals that are directly protected by vaccination.
	And we can calculate $\bar{\boldsymbol{\varsigma}}^{\mathrm{nf}}_{t}$ as:
	\begin{align}
		\bar{\boldsymbol{\varsigma}}^{\mathrm{nf}}_{t} = \eta\operatorname{diag}\left(\left(\tilde{\boldsymbol{\mathit{c}}} - \boldsymbol{\mathit{c}}^*(t)\right) \oslash\boldsymbol{s}^*(t)\odot\boldsymbol{\sigma}(t)\oslash\boldsymbol{\mathit{\rho}}\right), \label{eq:final_state_direct}
	\end{align}
	Furthermore, $\bar{\boldsymbol{\varsigma}}^{\mathrm{nf,c}}_{t} = \bar{\boldsymbol{\varsigma}}^{\mathrm{nf}}_{t}$, 
	$\bar{\boldsymbol{\varsigma}}^{\mathrm{nf,d}}_{t} = \bar{\boldsymbol{\varsigma}}^{\mathrm{nf}}_{t}\boldsymbol{\epsilon}$, 
	$\bar{\boldsymbol{\varsigma}}^{\mathrm{nf,y}}_{t} = \bar{\boldsymbol{\varsigma}}^{\mathrm{nf}}_{t}\left(\boldsymbol{\epsilon}\odot \boldsymbol{\phi}\right)$ correspond to the direct effects relevant for minimizing cumulative infections, deaths, and YLL, respectively.
	
	The total effects (comprising both direct and indirect components) can be characterized by the Jacobian matrix of the final state with respect to vaccine allocation. By the differentiation Eqs.~\ref{eq:vac_final_state}--\ref{eq:zeta_star}, the Jacobian matrix satisfies the following transcendental equation:
	\begin{align}
		\frac{\partial \tilde{\boldsymbol{\mathit{c}}}}{\partial\boldsymbol{\vartheta}_{t}} = - \operatorname{diag}\left(\boldsymbol{\sigma}(t)\right) + \operatorname{diag}\left(\boldsymbol{\zeta}^*(t)\odot\boldsymbol{\sigma}(t)\right) + k\lambda_{\mathrm{eff}}\operatorname{diag}\left(\left(\boldsymbol{s}^*(t) - \boldsymbol{\vartheta}^{\mathrm{eff}}_{t}\right)\odot\boldsymbol{\zeta}^*(t)\right)\boldsymbol{A}\operatorname{diag}\left(\boldsymbol{\rho}\right)	\frac{\partial \tilde{\boldsymbol{\mathit{c}}}}{\partial\boldsymbol{\vartheta}_{t}}.
		\label{eq:grad_c}
	\end{align}
	The indirect effects observable under the FS-DVP framework is denoted by $\check{\boldsymbol{\varsigma}}^{\mathrm{nf}}_{t}$ and could be calculated by subtracting the direct effects:
	\begin{align}
		\check{\boldsymbol{\varsigma}}^{\mathrm{nf}}_{t} = \left(\left.\frac{\partial \tilde{\boldsymbol{\mathit{c}}}}{\partial\boldsymbol{\vartheta}_{t}}\right|_{\boldsymbol{\vartheta}_t=\boldsymbol{0}}\operatorname{diag}\left(\boldsymbol{1} \oslash \boldsymbol{\mathit{\rho}}\right)\right)^{\mathsf{T}}
		- \bar{\boldsymbol{\varsigma}}^{\mathrm{nf}}_{t}, \label{eq:final_state_indirect}
	\end{align}
	where $\operatorname{diag}\left(\boldsymbol{1} \oslash \boldsymbol{\mathit{\rho}}\right)$ rescales the vaccine allocation such that each age group receives an equalized amount. 
	Finally, we define $\check{\boldsymbol{\varsigma}}^{\mathrm{nf,c}}_{t} = \check{\boldsymbol{\varsigma}}^{\mathrm{nf}}_{t}$, 
	$\check{\boldsymbol{\varsigma}}^{\mathrm{nf,d}}_{t} = \check{\boldsymbol{\varsigma}}^{\mathrm{nf}}_{t}\boldsymbol{\epsilon}$, 
	$\check{\boldsymbol{\varsigma}}^{\mathrm{nf,y}}_{t} = \check{\boldsymbol{\varsigma}}^{\mathrm{nf}}_{t}\left(\boldsymbol{\epsilon}\odot \boldsymbol{\phi}\right)$, 
	which correspond to the indirect effects relevant for minimizing cumulative infections, deaths, and YLL, respectively.
	
	We denote by $\bar{\boldsymbol{\varsigma}}^{\mathrm{tf}}_{t}$ the direct effects observable under the TF-DVP framework, the key idea behind its construction is to quantify, 
	for each vaccinated individual, the reduction in the incremental contribution to the objective within its own age group once the vaccine becomes effective. 
	This can be expressed as
	\begin{align}
		\bar{\boldsymbol{\varsigma}}^{\mathrm{tf}}_{t} = \eta\operatorname{diag}\left(\boldsymbol{\mathit{j}}^*(t)\odot\boldsymbol{\sigma}(t)\oslash\boldsymbol{\mathit{\rho}}\right), \label{eq:transient_state_direct}
	\end{align}
	We further define $\bar{\boldsymbol{\varsigma}}^{\mathrm{tf,c}}_{t} = \bar{\boldsymbol{\varsigma}}^{\mathrm{tf}}_{t}$, 
	$\bar{\boldsymbol{\varsigma}}^{\mathrm{tf,d}}_{t} = \bar{\boldsymbol{\varsigma}}^{\mathrm{tf}}_{t}\boldsymbol{\epsilon}$, 
	$\bar{\boldsymbol{\varsigma}}^{\mathrm{tf,y}}_{t} = \bar{\boldsymbol{\varsigma}}^{\mathrm{tf}}_{t}\left(\boldsymbol{\epsilon}\odot \boldsymbol{\phi}\right)$, 
	which correspond to the direct effects relevant for minimizing cumulative infections, deaths, and YLL, respectively.
	
	\subsection{Calculation of Marginal Vaccination Benefit}
	MVB quantifies the marginal reduction in an optimization objective, such as final cumulative infections, deaths, or YLL, achievable by administering an infinitesimal vaccine dose exclusively to a specific age group. 
	Accordingly, the MVB for each age group at time $t$ with respect to different objectives is defined via the Jacobian matrix of the final state with respect to vaccine allocation:
	\begin{align}
		\boldsymbol{\xi}^{\mathrm{c}}_{t} &= - \left(\left.\frac{\partial \tilde{\boldsymbol{\mathit{c}}}}{\partial\boldsymbol{\vartheta}_{t}}\right|_{\boldsymbol{\vartheta}_t=\boldsymbol{0}}\right)^{\mathsf{T}}\boldsymbol{\rho}\oslash \boldsymbol{\mathit{\rho}}, \\
		\boldsymbol{\xi}^{\mathrm{d}}_{t} &= - \left(\left.\frac{\partial \tilde{\boldsymbol{\mathit{c}}}}{\partial\boldsymbol{\vartheta}_{t}}\right|_{\boldsymbol{\vartheta}_t=\boldsymbol{0}}\right)^{\mathsf{T}}\left(\boldsymbol{\rho}\odot\boldsymbol{\epsilon}\right)\oslash \boldsymbol{\mathit{\rho}},\\
		\boldsymbol{\xi}^{\mathrm{y}}_{t} &= - \left(\left.\frac{\partial \tilde{\boldsymbol{\mathit{c}}}}{\partial\boldsymbol{\vartheta}_{t}}\right|_{\boldsymbol{\vartheta}_t=\boldsymbol{0}}\right)^{\mathsf{T}}\left(\boldsymbol{\rho}\odot\boldsymbol{\epsilon}\odot\boldsymbol{\phi}\right)\oslash \boldsymbol{\mathit{\rho}}, \label{eq:mvb}
	\end{align}
	$\boldsymbol{\xi}^{\mathrm{c}}_{t}$, $\boldsymbol{\xi}^{\mathrm{d}}_{t}$, and $\boldsymbol{\xi}^{\mathrm{y}}_{t}$ denote the MVB associated with reducing cumulative infections, deaths, and YLL, respectively.
	
	\subsection{Approach for Simulating Outbreaks across Different Countries}
	A standard approach for estimating the basic reproduction number ($R_0$) involves applying the Euler-Lotka equation~\cite{park2021forward,wallinga2007generation,dietz1993estimation}: $R_0 = 1/\int_{0}^{+\infty} {e^{-g\tau} \psi_{\mathrm{gen}}(\tau) d\tau}$,
	where $g$ represents the growth rate and it is the exponential rate at which the cumulative number of infections increases, assuming an exponential growth pattern at the early stage of an outbreak,
	and $\psi_{\mathrm{gen}}(\tau)$ represents the generation time distribution, satisfying $\psi_{\mathrm{gen}}(\tau) = \omega_{\mathrm{inf}}(\tau)\Psi_{\mathrm{rem}}(\tau)$.
	
	We fit the early epidemic curve, represented by cumulative infection data, to an exponential growing model, enabling estimation of $g$ and its 95\% confidence interval. 
	The generation time distribution, $\psi_{\mathrm{gen}}(\tau)$, for COVID-19 is well described by a Weibull distribution parameterized by shape ($\alpha_{\mathrm{gen}}$) and scale ($\beta_{\mathrm{gen}}$) parameters, 
	each with associated 95\% confidence intervals~\cite{ferretti2020quantifying}. To account for uncertainty, we are able to independently sample 1000 random sets of $g$, $\alpha_{\mathrm{gen}}$, and $\beta_{\mathrm{gen}}$, 
	and compute the corresponding 1000 $R_0$ values, yielding 1000 estimates for the parameter $k$.
	
	For each country, using country-specific epidemic data, we generate 1000 sets of dynamic parameters ($k$, $\alpha_{\mathrm{gen}}$, and $\beta_{\mathrm{gen}}$). We then simulate disease transmission 1000 times independently for each country. For one country, throughout these 1000 simulations, parameters such as population distribution, contact matrices, IFRs, immune response delays, and life expectancy are held fixed, based on country-specific data sources~\cite{WorldPop64:online,prem2017projecting,ferretti2020quantifying,osmc,omori2020age,Actualiz44:online,GHOBycat60:online}.

	\section*{Data Availability}
	\vspace*{-0.1in}
	All the data used in this study are available from the corresponding author upon request.
	
	\section*{Code Availability}
	\vspace*{-0.1in}
	The GitHub repository which includes the source code for all the figure results can be accessed at
	{\color{blue} \url{https://github.com/fengmi9312/Final-State-Dynamic-Vaccine-Prioritization.git}}.
	
	\section*{Acknowledgments}
	\vspace*{-0.1in}
	This work was supported by the Hong Kong Baptist University (HKBU) Strategic Development Fund. 
	This research was conducted using the resources of the High-Performance Computing Cluster Centre at HKBU, 
	which receives funding from the Hong Kong Research Grant Council and the HKBU.
	
	\section*{Author Contributions}
	\vspace*{-0.1in}
	M.F., L.T. and C.-S.Z. designed research; M.F. performed research; L.T. and C.-S.Z. contributed analytic tools; M.F., L.T. and C.-S.Z. analysed data; M.F., L.T. and C.-S.Z. discussed the results and wrote the paper.
	
	\section*{Competing Interests}
	\vspace*{-0.1in}
	The authors declare no competing interests.
	
	\section*{Correspondence}
	\vspace*{-0.1in}
	To whom correspondence should be addressed: liangtian@hkbu.edu.hk, cszhou@hkbu.edu.hk
	
	\bibliographystyle{naturemag}

	\end{document}